\documentclass{article}

\usepackage{arxiv}

\usepackage[utf8]{inputenc} 
\DeclareUnicodeCharacter{03C9}{$\omega$} 
\usepackage[T1]{fontenc}    
\usepackage{hyperref}       
\usepackage{url}            
\usepackage{booktabs}       
\usepackage{amsfonts}       
\usepackage{nicefrac}       
\usepackage{microtype}      
\usepackage{lipsum}

\usepackage{cite}
\usepackage{amsmath,amssymb,amsfonts}
\usepackage{amsthm}
\usepackage{graphicx}
\usepackage{algorithm}
\usepackage[noend]{algpseudocode}
\usepackage{hyperref}
\usepackage{textcomp}
\usepackage{threeparttablex} 

\usepackage{mathtools}
\usepackage{caption}
\usepackage{float}
\usepackage{stmaryrd}
\usepackage{epstopdf}
\usepackage{multirow}
\usepackage{pifont}
\usepackage{caption}
\usepackage{subcaption}
\usepackage{xcolor}
\usepackage{tabularx}

\newcommand{\R}{{\mathbb{R}}}

\newcommand{\B}{{\mathcal B}}

\newcommand{\N}{{\mathbb{N}}}

\newcommand{\V}{{\mathcal{V}}}

\newcommand{\X}{{\mathbf{X}}}

\newcommand{\tx}{{t_{\times}}}

\newcommand{\always}{\Box}
\newcommand{\eventually}{\Diamond}

\newcommand{\RD}[1]{#1}

\newtheorem{thm}{Theorem}[section]

\newtheorem{assum}{Assumption}

\newtheorem{defn}[thm]{Definition}

\newtheorem{rem}[thm]{Remark}

\newenvironment{pf}{\par\noindent\textbf{Proof.}\ }{\hfill$\square$\par}

\newtheorem{example}{Example}

\newenvironment{contexample}[1]{%
  \renewcommand{\theexample}{\ref{#1} (continued)}%
  \begin{example}%
  \def\@currentlabel{\ref{#1}}%
}{%
  \end{example}%
}

\title{Scalable and Approximation-free Symbolic Control for Unknown Euler–Lagrange Systems
\thanks{ This work was supported in part by the ARTPARK. The work of Ratnangshu Das was supported by the Prime Minister’s Research Fellowship from the Ministry of Education, Government of India.}
}

\author{
 Ratnangshu Das \\
  Centre for Cyber-Physical Systems\\
  IISc, Bengaluru, India\\
  \texttt{ratnangshud@iisc.ac.in} \\
   \And
 Shubham Sawarkar \\
 Centre for Cyber-Physical Systems\\
  IISc, Bengaluru, India\\
  \texttt{shubhamsg@iisc.ac.in} \\
  \And
 Pushpak Jagtap \\
  Centre for Cyber-Physical Systems\\
  IISc, Bengaluru, India\\
  \texttt{pushpak@iisc.ac.in} \\
}

\begin{document}
\maketitle

\begin{abstract}
We propose a novel symbolic control framework for enforcing temporal logic specifications in Euler–Lagrange systems that addresses the key limitations of traditional abstraction-based approaches. Unlike existing methods that require exact system models and provide guarantees only at discrete sampling instants, our approach relies only on bounds on system parameters and input constraints, and ensures correctness for the full continuous-time trajectory. The framework combines scalable abstraction of a simplified virtual system with a closed-form, model-free controller that guarantees trajectories satisfy the original specification while respecting input bounds and remaining robust to unknown but bounded disturbances. We provide feasibility conditions for the construction of confinement regions and analyze the trade-off between efficiency and conservatism. Case studies on pendulum dynamics, a two-link manipulator, and multi-agent systems, including hardware experiments, demonstrate that the proposed approach ensures both correctness and safety while significantly reducing computation time and memory requirements. These results highlight its scalability and practicality for real-world robotic systems where precise models are unavailable and continuous-time guarantees are essential.
\end{abstract}

\section{Introduction}

The increasing use of autonomous systems in safety-critical domains has raised the demand for controllers that can guarantee satisfaction of complex tasks. Classical objectives, such as stabilization or trajectory tracking, are often insufficient, as many tasks involve logical and temporal structures; for example, a robot may need to visit regions or avoid unsafe areas in a specific temporal order. Temporal logics, such as Linear Temporal Logic (LTL) and related formalisms \cite{baier2008principles}, provide an efficient and formal way to express such specifications. Their use has improved the modeling of autonomy and the verification of controllers. Successful applications range from motion planning for mobile robots \cite{LaValle2006}, \cite{fainekos2009temporal}, \cite{kress2009temporal}, to coordination of multi-agent systems \cite{Kloetzer}, \cite{sundarsingh2023scalable}, and control of cyber–physical systems \cite{Belta2007}. Overall, temporal logic has enabled controller synthesis beyond low-level stabilization, providing formal guarantees of task satisfaction.

An interesting methodology that has gained significant attention in this context is the \textit{abstraction-based} controller synthesis techniques or the so-called \textit{symbolic control} \cite{tabuada2009verification}. The idea is to approximate a nonlinear dynamical system with a finite transition system through state and input discretizations, often called symbolic models, that preserve essential behavioral relations \cite{Alur2000}, \cite{tabuada2009verification}. Discrete synthesis is then carried out based on this abstraction and desired specifications using automata-theoretic tools, to obtain a discrete strategy. Finally, the resulting strategy is refined into a feedback controller that ensures the original system meets the specifications. 

Symbolic control has emerged as a powerful framework, offering formal guarantees and broad applicability. Symbolic abstractions have been developed for diverse classes of systems, including linear \cite{tabuada2006linear}, \cite{rungger2013specification}, \cite{yordanov2011temporal}, and nonlinear dynamics \cite{reissig2011computing}, \cite{zamani2011symbolic}, \cite{girard2012controller}. The range of specifications addressed spans from reach-avoid and safety requirements \cite{tabuada2009verification}, \cite{jagtap2020software} to full LTL specifications \cite{maler1995synthesis}, \cite{SCOTS}. The framework has also been extended to handle disturbances \cite{liu2016finite}, uncertain models \cite{liu2014abstraction}, and even stochastic dynamics \cite{abate2007computational}, \cite{jagtap2020symbolic}, \cite{zamani2014symbolic}, \cite{lavaei2022automated}. Furthermore, the approach is supported by various software tools such as OmegaThreads \cite{khaled2021omegathreads}, QUEST \cite{jagtap2017quest}, LTLMoP \cite{finucane2010ltlmop}, and SCOTS \cite{SCOTS}, enabling applications across diverse domains. These results demonstrate that symbolic control provides a unifying methodology for bridging formal task specifications and controllers.

Despite its strengths, the practical use of symbolic control faces three main challenges. \RD{The first is scalability. Constructing a symbolic abstraction typically requires discretizing the state and input spaces, resulting in a model that grows exponentially with system dimension.} Several enhancements, such as compositional abstraction techniques \cite{saoud2019compositional}, \cite{saoud2021compositional}, \cite{zamani2017compositional}, leveraging the incremental stability property \cite{girard2014approximately}, \cite{zamani2017towards}, the integration of barrier certificates \cite{sundarsingh2023scalable}, and incremental tree-based approximations \cite{Sampling_MAS}, have improved on the traditional abstraction-based method. However, the curse of dimensionality remains, making synthesis infeasible for high-dimensional systems. \RD{The second limitation is the reliance on accurate system models. In reality, physical systems are subject to modeling uncertainties and external disturbances, making the derivation of precise models challenging.} To address this, several works have introduced data-driven learning of approximate models \cite{makdesi2021efficient}, \cite{kazemi2024data}, \cite{nazeri2025data}. While these methods improve reliability under uncertainty, they can often be data-intensive and limited to low-dimensional systems. \RD{The third challenge is that most symbolic methods provide guarantees only at sampling instants, not over the full continuous-time trajectories.} Extensions that account for inter-sample behavior using continuous-time reachability \cite{saoud2018contract} or robustness margins \cite{liu2016finite}, \cite{liu2014abstraction}, address this gap, but are again dependent on accurate system knowledge.

\RD{In this paper, we propose a new framework to address the limitations of traditional symbolic control for Euler–Lagrange systems under temporal logic specifications. The approach introduces Virtual Confinement Zones (VCZs), an $n$-dimensional balls in the configuration space that act as a moving safe region for the system. Instead of synthesizing controllers for the original nonlinear dynamics, abstraction-based synthesis is carried out on a much simpler virtual system. The resulting controller is then systematically transferred to the original EL system in a model-free and computationally efficient manner. The key contributions of this paper are summarized as follows:
\begin{enumerate}
    \item In Section \ref{sec:vcz}, we introduce the notion of a VCZ and model the motion of its center using simple single-integrator dynamics.
    \item In Section \ref{sec:symbolic}, we reformulate the given temporal logic specification based on the VCZ radius and synthesize abstraction-based symbolic controllers for the VCZ center dynamics. Owing to the simplicity of the virtual system, the synthesis naturally becomes computationally efficient and scales well with system dimension.
    \item Next, in Section \ref{Sec:Contorl_for_VCZ}, we develop a closed-form, model-free controller that guarantees the actual system remains inside the VCZ at all times. This controller relies only on known bounds on system parameters and disturbances, and ensures satisfaction of the original specification for the real system, with guarantees for continuous-time trajectories.
    \item Finally, in Section \ref{Sec:Case Study}, we validate the proposed framework through case studies on a pendulum system, a two-link manipulator, and a multi-agent setup, showing that the computational advantages become more pronounced as the system dimension increases. 
\end{enumerate}
Overall, the proposed framework enables practical deployment of formally verified temporal-logic-based control on high-dimensional robotic systems, as further demonstrated by hardware experiments on an 8-dimensional multi-agent setup.}

\section{Preliminaries and Problem Formulation}
\label{sec:prelim}
\subsection{Notation}
The symbols $\N$, $\N_0$, $\R$, $\R^+$, and $\R_0^+ $ denote the set of natural, whole, real, positive real, and nonnegative real numbers, respectively. 
For $a,b\in\N$ and $a\leq b$, we use $[a;b]$ to denote a close interval in $\N$. 
For $x, y \in \R^n$, the vector inequalities, $x \preceq y$ (and $x \succeq y$) represents $x_i \leq y_i$ (and $x_i \geq y_i$), $\forall i \in [1;n]$.
We use $\mathbb{I}_n$ and $\textbf{0}_{n\times m}$ to denote identity matrix in $\R^{n\times n}$ and zero matrix in $\R^{n\times m}$, respectively. 
$x \uparrow (\downarrow) \ a$ indicates $x$ approaches $a$ from the left (right) side. 
Given a matrix $M\in\R^{n\times m}$, $M^\top$ represents the transpose of matrix $M$. 
Given $N \in \N$ sets $\X_i$, $i\in\left[1;N\right]$, we denote the Cartesian product of the sets by $\X=\prod_{i\in\left[1;N\right]}\X_i:=\{(x_1,\ldots,x_N)|x_i\in \X_i,i\in\left[1;N\right]\}$. 
For $a,b \in (\R \cup \{-\infty, \infty\})^n, a\preceq b$, the closed hyper-interval is denoted by $\llbracket a,b \rrbracket := \R^n \cap ([a_1,b_1] \times [a_2,b_2] \times \ldots \times [a_n,b_n])$.
We identify the relation $R \subseteq A \times B$ with the map $R: A \rightarrow 2^B$ defined by $b \in R(a)$ iff $(a,b) \in R$. Given a relation $R \subseteq A \times B$, $R^{-1} := \{(b,a) \in B \times A \mid (a,b) \in R\}$. The composition of two maps $Q$ and $R$ is $Q\circ R(x):=Q(R(x))$. The map $R$ is said to be strict when $R(a) \neq \emptyset$ for every $a \in A$.
A ball centered at $\xi \in \R^n$ with radius $\lambda \in \R^+$ is defined as $\mathcal{B}(\xi, \lambda) := \{ x \in \R^n \mid \|x - \xi\| \leq \lambda \}$.
All other notation in this paper follows standard mathematical conventions.

\subsection{System Definition}
We consider \RD{a fully-actuated Euler-Lagrange (EL) system} $\mathcal{S}$ described by the dynamics:
\begin{align}
    \mathcal{S}: M(x)\Ddot{x} + V(x,\dot{x}) + G(x) = \tau + d(t), \label{eqn:sysdyn}
\end{align}
where $x(t) = [x_1(t), \ldots, x_n(t)]^\top \in X \subset \R^n$ represents the generalized coordinates or the system configuration, $\dot{x}(t)$ is the system velocity, $\tau(t) \in \R^n$ is the control input, and $d(t) \in \mathbb{D} \subset \R^n$ is an unknown bounded external disturbance. 
The terms $M(x) \in \R^{n \times n}$, $V(x, \dot{x}) \in \R^n$, and $G(x) \in \R^n$ correspond to the mass matrix, Coriolis and centrifugal forces, and gravitational effects, respectively. For the reader’s convenience, we simplify the notation by omitting arguments and parentheses when it is clear that a symbol represents a function. For example, $M(x), V(x,\dot{x}), G(x)$ and $d(t)$ are denoted as $M,V,G$ and $d$, respectively. 

\begin{assum} \label{assum:unknown dynamics}
    The mass matrix $M(x)$, the Coriolis and centrifugal terms $V(x, \dot{x})$, the gravity vector $G(x)$ and the external disturbance $d(t)$ are all unknown.
\end{assum}

For the EL system $\mathcal{S}$, consider the velocity bound $\overline{v} \in \R^n$ and a bounded control input with a bound $\overline{\tau} \in \R^n$ such that $|\dot{x}(t)| \preceq \overline{v}$ and $|\tau(t)| \preceq \overline{\tau}$ for all $t \in \R_0^+$. Under this condition, the corresponding bounds on the norms of the system parameters are available \cite[Chapter 2]{ELbook}, \cite[Chapter 7]{ELbook2}, \cite{ELbounds}. Although the system parameters $M(x)$, $V(x,\dot{x})$, $G(x)$, and the disturbance $d(t)$ are unknown, their boundedness can be used in control design. Assumption~\ref{assum:sys_bounds} summarizes these bounds, which will be used to establish the feasibility conditions in Section~\ref{sec:feas}.

\begin{assum}\label{assum:sys_bounds}
For a system dynamics in \eqref{eqn:sysdyn}, the external disturbance $d$ is bounded as $-\overline{d} \preceq d(t) \preceq \overline{d}$ for all $t \in \R_0^+$ with some $\overline{d}\in\R^n$, and under the scaling of $M^{-1}$ it satisfies $-\underline{m}_i \overline{d} \preceq M^{-1}d \preceq \underline{m}_i \overline{d}$ for some $\underline{m}_i \in \R^+$. Given the control limit $\overline{\tau}$, there exists a constant $\underline{m} \in \R^+$ such that $\underline{m}\overline{\tau} \preceq M^{-1}\overline{\tau}$. Coriolis and gravitational terms $V_M := -M^{-1}(V+G)$ are bounded as $\underline{V}_M \preceq V_M \preceq \overline{V}_M$, with $\underline{V}_M, \overline{V}_M \in \R^n$ with $V_M^{\max}:=\max(-\underline{V_M}, \overline{V_M})$. 
\end{assum}

\subsection{Problem Formulation and Schematic of Solution} 
We consider the EL system $\mathcal{S}$ described by \eqref{eqn:sysdyn}, subject to Assumptions~\ref{assum:unknown dynamics} and \ref{assum:sys_bounds}. Our goal is to design a model-free control strategy that guarantees satisfaction of a given temporal logic specification $\phi$ (usually expressed as Linear temporal logic (LTL) or (in)finite sequence over automata \cite{baier2008principles}). The proposed controller does not require exact knowledge of the system dynamics and relies only on known bounds on system parameters. It is robust to unknown but bounded disturbances and respects input constraints, ensuring that all closed-loop trajectories of $\mathcal{S}$ satisfy the prescribed specification.

\RD{To make the proposed approach easier to follow, we introduce a running example~\ref{Running Example} based on a classical pendulum system. This example is used throughout the paper to illustrate each step of the method and to provide intuition for the design choices.}

\begin{example}\label{Running Example}
Consider the pendulum dynamics given by
\begin{equation}\label{eqn:pend}
    \frac{ml^2}{3}\Ddot{x} + \frac{mgl}{2}\sin(x) = \tau + d,
\end{equation}
where $x$ and $\dot{x}$ are the angular position and velocity of the pendulum, $\tau$ is the control torque input, and $d$ is an unknown but bounded disturbance. The parameters $m, g$, and $l$ correspond to the pendulum mass, acceleration due to gravity, and rod length, respectively. The input constraint is given by $\tau \in [-2,2]$~Nm ($\overline{\tau} = 2$~Nm). In addition, the velocity bound is given by $\dot{x} \in [-0.1,0.1]$~rad/s ($\overline{v} = 0.1$~rad/s) with the configuration space is restricted to $X = [-0.2, 0.2]$~rad.\\
Given the operating region and input constraints, the system parameters lie within the following bounds:
$\underline{m} = \underline{m}_i = 3 \text{ kg}^{-1}\text{m}^{-2}, V_M^{\max} = 1 \text{ Nm}, \overline{d} = 0.5 \text{ Nm}$.\\
This setup represents a nonlinear EL system with unknown parameters and bounded disturbances, making it ideal for demonstrating the effectiveness of our proposed control framework.\qed
\end{example}

\begin{figure*}[t]
    \centering
    \includegraphics[width=\textwidth]{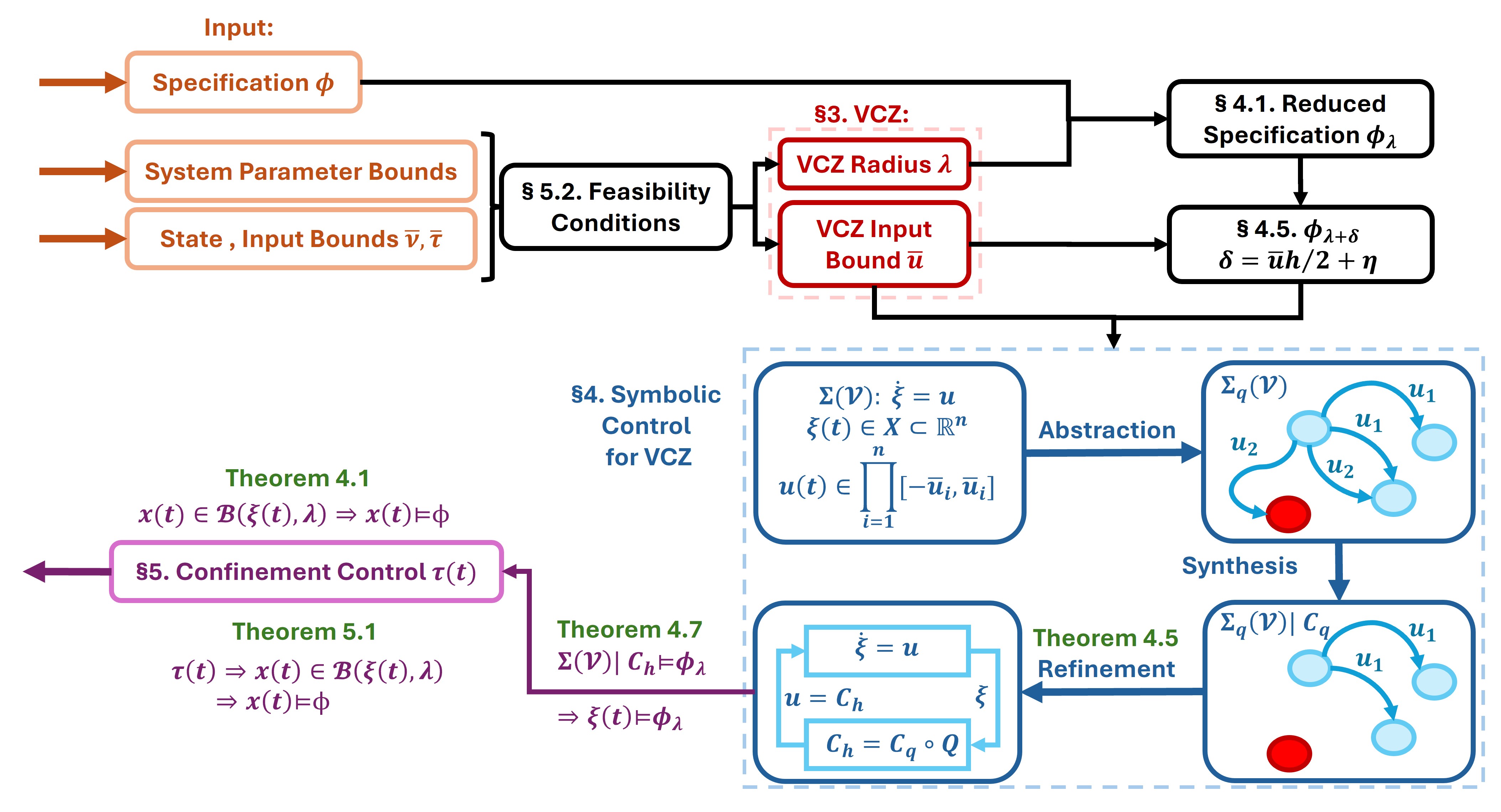}
    \caption{Schematic of the proposed framework.}
    \label{fig:framework}
\end{figure*}

\RD{The key intuition behind our solution is to separate the problem into two layers.\\
$(i)$ First, we introduce a Virtual Confinement Zone (VCZ), which is an $n$-dimensional ball whose center follows simple first-order dynamics. Using known bounds on system parameters and inputs, we determine a feasible VCZ radius. We then modify the original temporal logic specification to account for this radius and synthesize a symbolic controller for the VCZ center. This step is computationally efficient because it involves only a low-dimensional, single-integrator system.\\
$(ii)$ Second, we design a closed-form, approximation-free controller that guarantees the actual system remains inside the VCZ for all time. As long as the system stays inside the VCZ, satisfaction of the modified specification by the VCZ center implies satisfaction of the original specification by the real system.\\
The overall flow of the framework is shown in Figure~\ref{fig:framework}.}

\section{Virtual Confinement Zone}\label{sec:vcz}
We introduce the notion of a Virtual Confinement Zone (VCZ) as a geometric ball that constrains the evolution of the system configuration within a prescribed region of the configuration space. Consider the EL system $\mathcal{S}$ in \eqref{eqn:sysdyn}, with $x(t) \in X \subset \R^n$. A VCZ is defined as an $n$-dimensional closed ball $\B(\xi(t), \lambda) \subset X$, centered at $\xi(t) \in X$ with radius $\lambda \in \R^+$. Intuitively, the VCZ can be seen as a moving “safe zone” that encapsulates all admissible trajectories of the system, with its radius $\lambda$ capturing robustness to uncertainties.\\
The temporal evolution of the VCZ is determined by the dynamics of its center $\xi(t)$. We model these dynamics as a first-order control-affine system:
\begin{equation}\label{eqn:vcz}
    \V : \dot{\xi}(t) = u(t), \ u(t) \in U \subset \R^n.     
\end{equation}
where $u(t)$ denotes the virtual input that governs the trajectory of the VCZ center. 
The input set $U$ is taken to be an axis-aligned hyperrectangle with componentwise bounds, $\overline{u} = \Big[ \overline{u}_1,\ldots, \overline{u}_n \Big]^\top \in \mathbb{R}^n$, such that
\begin{equation}\label{eqn:U-box}
U := \{ u \in \R^n \mid -\overline{u} \preceq u \preceq \overline{u} \}
= \prod_{i=1}^n \big[-\overline{u}_i, \overline{u}_i\big].
\end{equation}

The radius $\lambda$ and the input bound $\bar{u}$ are design parameters, which we assume are given for now. A detailed discussion on feasibility conditions and systematic selection of appropriate $\lambda$ and $\bar{u}$ is provided in Section \ref{sec:lam-ub}.

\begin{contexample}{Running Example}
    For pendulum dynamics in \eqref{eqn:pend}, we define the VCZ as the ball $\B(\xi(t), \lambda)$ centered at $\xi(t) \in \R$, with radius $\lambda = 0.018$~rad. \qed
\end{contexample}

\begin{defn}[Reachable Set for VCZ Dynamics]
For the VCZ system in \eqref{eqn:vcz}, the reachable set at time $h \ge 0$ from the initial hyperrectangle $X_0 = \prod_{i=1}^{n} [\underline{x}_{0,i}, \overline{x}_{0,i}]$, under input $u = [u_1, \ldots, u_n]^\top \in U$ is given exactly by
$\mathsf{Reach}_h(X_0, u) = \prod_{i=1}^{n} [\underline{x}_{0,i}+hu_{i}, \overline{x}_{0,i}+hu_{i}].$
\end{defn}

\section{Symbolic Control for VCZ System}\label{sec:symbolic}

In this section, we first formalize how the original system specification is modified using the VCZ radius $\lambda$. We then discuss the role of symbolic control in this setting, and demonstrate how it allows the systematic synthesis of controllers to guarantee that the VCZ system $\mathcal{V}$ in \eqref{eqn:vcz} satisfy this modified specification.

\subsection{Modification of Specification} \label{sec:reduce_spec}
\RD{The intuition behind the modification of specification is that the VCZ represents a finite region of the configuration space, characterized by its radius $\lambda$, within which the original system's trajectory $x:\R_0^+ \rightarrow X \subset \R^n$ is constrained to evolve $\|x(t)-\xi(t)\| \leq \lambda$ for all $t\in\R_0^+$. Consequently, if the original system is required to satisfy a specification $\phi$, i.e., $x \models \phi$, then it suffices to require that the VCZ center trajectory $\xi: \R_0^+ \rightarrow X \subset \R^n$ satisfies a modified specification $\phi_\lambda$, i.e., $\xi \models \phi_\lambda$.}
In the following, we first detail this modification procedure for reach-avoid-stay specifications, and subsequently extend it to general temporal logic tasks.

\subsubsection{Reach-Avoid-Stay (RAS) Specification}
Consider a reach-avoid-stay (RAS) specification $\phi$ defined for the system $\mathcal{S}$, and a fixed VCZ radius $\lambda \in \R^+$. The objective is to construct a modified specification $\phi_\lambda$ that represents a modified version of the original task. 

The RAS specification is expressed as an LTL Formula:  
$\phi := \eventually G \wedge \always (X \setminus O),$
where $\eventually$ and $\always$ denote the temporal operators \textit{eventually} and \textit{always} \cite{baier2008principles}. It requires that the trajectory eventually reach the goal set $G \subseteq X$, never enter the unsafe set $O \subseteq X$, and always remain within the space $X$.

For a given VCZ radius $\lambda$, the modified specification $\phi_\lambda$ is obtained by modifying the original specification $\phi$: the goal set, and configuration space are shrunk by $\lambda$, while the unsafe set is expanded by $\lambda$, resulting in
\begin{align*}
    G_\lambda &:= \{ z \in G \mid z' \in G, \forall z' \text{ with } \|z - z'\| \leq \lambda \}, \\   
    O_\lambda &:= \{ z \in \R^n \mid \|z - z'\| \leq \lambda, \forall z' \in O \}, \\
    X_\lambda &:= \{ z \in X \mid z' \in X, \forall z' \text{ with } \|z - z'\| \leq \lambda \}.
\end{align*}
The corresponding modified reach-avoid-stay specification is therefore defined as:
\begin{align}\label{eqn:redspec}
    \phi_\lambda := \eventually G_\lambda \wedge \always (X_\lambda \setminus O_\lambda). 
\end{align}
\begin{rem}
\RD{The choice of the VCZ radius $\lambda$ directly affects the modified specification in \eqref{eqn:redspec}. A larger $\lambda$ leads to a more conservative modification, and in highly cluttered environments the modified specification can become infeasible. For this reason, $\lambda$ should generally be chosen as small as possible to limit conservatism. A detailed discussion on selecting $\lambda$ is provided in Section~\ref{sec:lam-ub}.}
\end{rem}

\subsubsection{Temporal Logic Specifications}
We now extend this idea of modification of specifications from RAS tasks to enforce more complex specifications formulated, for example, in linear temporal logic (LTL) \cite{LTL} or linear temporal logic over finite traces (LTL$_F$) \cite{de2013linear} or ($\omega$-)regular automata \cite{baier2008principles}. 

LTL specifications over continuous state spaces can be systematically decomposed into a finite sequence of RAS sub-tasks by leveraging automata-based representation of specifications (For more details, refer to \cite{NAHS}, \cite{majumdarregret_Decomp_Spec}, \cite{cai2022learning_Decomp_Spec}, \cite{jackermeier2024deepltl_Decomp_Spec}). Specifically, an LTL formula $\phi$ can be expressed as an ordered sequence of $N_s \in \mathbb{N}$ sub-tasks: $\phi = \phi^1 \rightarrow \phi^2 \rightarrow \ldots \rightarrow \phi^{N_s}$, where each sub-task $\phi^i$ takes the form of RAS task as:
$\phi^i := \eventually G^i \wedge \always (X \setminus O^i).$
This requires the trajectory $\xi(t)$ to eventually reach the goal set $G^i$, always avoid the unsafe set $O^i$, and always remain within the admissible space $X$.

This sequential decomposition is particularly powerful, as it breaks down complex temporal specifications into simpler RAS primitives. Thus, the overall specification can be systematically modified by adjusting each sub-task with parameter $\lambda$. In other words, by shrinking the admissible space and target sets to $X_\lambda$ and $G^i_\lambda$, respectively, and expanding the unsafe sets to $O^i_\lambda$, we obtain a modified specification $\phi_\lambda = \phi^1_\lambda \rightarrow \phi^2_\lambda \rightarrow \ldots \rightarrow \phi^{N_s}_\lambda$, with $\phi^i_\lambda := \eventually G^i_\lambda \wedge \always (X_\lambda \setminus O^i_\lambda)$.

This modification specifies a task for the center trajectory $\xi$ of the VCZ $\mathcal{B}(\xi(t), \lambda)$.  If the configuration of the original system remains within the VCZ, then the original system is guaranteed to satisfy the specification $\phi$, provided that $\xi$ satisfies the modified specification $\phi_\lambda$.

\begin{thm}\label{thm:red}
    Let the system $\mathcal{S}$ in \eqref{eqn:sysdyn} be assigned an LTL specification $\phi$. Consider a Virtual Confinement Zone (VCZ) as the ball $\mathcal{B}(\xi(t), \lambda)$, where the center trajectory $\xi: \R_0^+ \rightarrow X \subset \R^n$ evolves under the dynamics in Equation~\eqref{eqn:vcz}, and $\lambda > 0$ is the fixed radius.     
    If $\xi$ satisfies the modified specification $\phi_\lambda$, and the system configuration remains confined within the VCZ for all time, then the system trajectory $x:\R_0^+ \rightarrow X \subset \R^n$ satisfies the original specification $\phi$:
    \begin{align}\label{eqn:red_satisfy}
        \xi \models \phi_\lambda \text{ and } \|x(t) - \xi(t)\| < \lambda \implies x \models \phi.
    \end{align}
\end{thm}

\begin{pf}
    See Appendix \ref{sec:red_proof}
\end{pf}

\begin{contexample}{Running Example}
    Returning to the pendulum example, we consider the invariance specification:
    $\phi = \always X,$
    with $X=[-0.2, 0.2]$, i.e., the pendulum’s angle should \textit{always} remain within a small neighborhood around the upright position.\\ 
    By selecting a VCZ radius $\lambda = 0.018$~rad, the modified specification  becomes $\phi_\lambda = \always X_\lambda,$ with $X_\lambda=[-0.182,0.182]$.\\
    This modification guarantees that if the pendulum trajectory $x(t)$ remains confined within the VCZ around $\xi(t)$, then the original specification $\phi$ is satisfied (as discussed in Theorem \ref{thm:red}). \qed
\end{contexample}

\subsection{Symbolic Control Fundamentals}
We now provide a systematic approach to design the symbolic controller ensuring the modified specification $\phi_\lambda$. The synthesis approach uses the symbolic model of the VCZ system $\V$ in Equation~\eqref{eqn:vcz}. The advantage of using this approach is that it provides a correct-by-construction controller for high-level specifications \cite{tabuada2009verification} for nonlinear control systems. Apart from the ability to handle nonlinear systems, it efficiently handles constraints on state and input space and provides maximally permissible controllers.

\subsubsection{Transition Systems and Equivalence Relation}
We begin by introducing the notion of transition systems \cite{tabuada2009verification}, which will serve as a representation for the VCZ control system and its corresponding symbolic models.
\begin{defn}[Transition System \cite{tabuada2009verification}]
    A transition system is a tuple $\Sigma = (X, X_0, U, \rightarrow)$, where $X$ is the set of states, $X_0 \subseteq X$ is the set of initial states, $U$ is the set of inputs, and $\rightarrow \subseteq X \times U \times X$ is the transition relation.
\end{defn}

A transition $(x,u,x') \in \rightarrow$ is alternatively written as $x \xrightarrow{u} x'$, indicating that under input $u \in U$, the system evolves from state $x$ to the successor state $x'$. The set of all $u$-successors of $x$ is denoted by $Post_u(x)$, while $U(x)$ represents the set of admissible inputs for which $Post_u(x)$ is non-empty. 

To connect the continuous VCZ control system $\V$ in Equation~\eqref{eqn:vcz} and its symbolic model, we use the concept of feedback refinement relation (FRR) \cite{Reissig2017feedback}, which is later used in symbolic controller synthesis for $\V$.

\begin{defn}[Feedback Refinement Relation \cite{Reissig2017feedback}]\label{Def:FRR}
Let $\Sigma_1 = (X_1, X_{0,1}, U_1, \xrightarrow[1]{})$ and $\Sigma_2 = (X_2, X_{0,2}, U_2, \xrightarrow[2]{})$ be two transition systems. A strict relation $Q \subseteq X_1 \times X_2$ is called a feedback refinement relation from $\Sigma_1$ to $\Sigma_2$, denoted $\Sigma_1 \preceq_Q \Sigma_2$, if for every $(x_1, x_2) \in Q$:
\begin{itemize}
    \item 
    $U_2(x_2) \subseteq U_1(x_1)$, and
    \item 
    $Q(Post_u(x_1)) \subseteq Post_u(x_2)$.
\end{itemize}
\end{defn}
Therefore, an FRR $Q$ allows a controller designed for an abstract system $\Sigma_2$ to be refined into a valid controller for the original system $\Sigma_1$. Further details about FRR and its role in controller synthesis can be found in \cite{Reissig2017feedback}.

\subsubsection{VCZ System as a Transition System}

The transition system associated with the VCZ system $\V$ in \eqref{eqn:vcz} with a sampling time $h$ is given by the tuple
$\Sigma_h (\V) = ( X_h, X_{0,h}, U_h, \xrightarrow[h]{} ),$
where $X_h = X, X_{0,h} = X_0, U_h = U$, and $x \xrightarrow[h]{u} x'$ is a transition if and only if there exists $x' = x +uh$, where $u \in U_h$. Note that we abuse the notation above by identifying $u$ with the constant input curve with domain $[0,h]$ and value $u$.

\begin{defn}[Controlled System]
Consider a transition system $\Sigma_h(\V) = (X_h, X_{0,h}, U_h, \xrightarrow[h]{})$ and a memoryless controller $C_h : X_h \rightrightarrows U_h$, where for all $x_h \in X_h$, $C_h(x_h) \subseteq U_h(x_h)$. The domain of $C_h$ is defined as $\mathrm{dom}(C_h) := \{ x_h \in X_h \mid C_h(x_h) \neq \emptyset \}$. The controlled system $\Sigma_h(\V) \mid C_h$ is then given by the tuple 
$\Sigma_h(\V) \mid C_h := (X_{C,h}, X_{0,C,h}, U_{C,h}, \xrightarrow[C,h]{}),$
where $X_{C,h} = X_h \cap \text{dom}(C_h)$, $X_{0,C,h} \subseteq X_{C,h}$, and $U_{C,h} = U_h$. For any $x_{C,h} \in X_{C,h}$, and $u_{C,h} \in U_{C,h}$, the transition $x_{C,h}' \in Post_{u_{C,h}}(x_{C,h})$ holds iff $u_{C,h} \in C_h(x_{C,h})$.
\end{defn}

\subsubsection{Symbolic Model for VCZ System}
To design controllers for the concrete system $\Sigma_h(\V)$ from its symbolic model, the system and its symbolic model must satisfy formal behavioral inclusions in terms of FRR, as in Definition~\ref{Def:FRR}. Consider the sampling times $h \in \R^+$ and quantization parameter $\eta \in (\R^+)^n$. The symbolic model of $\Sigma_h(\V)$ is given by the tuple
$\Sigma_q(\V) = (X_q, X_{0,q}, U_q, \xrightarrow[q]{})$,
where 
\begin{itemize}
    \item $X_q$ is a cover of $X_h$ consisting of non-empty closed hyperintervals called cells. For computation of the symbolic model, we consider a compact subset $\overline{X}_q \subset X_q$ of congruent hyperrectangles aligned on a uniform grid parametrized with a quantization parameter $\eta \in (\R^+)^n$. Each cell $x_q \in \overline{X}_q$ is represented as $x_q = c_{x_q} + \left \llbracket -\frac{\eta}{2}, \frac{\eta}{2} \right \rrbracket$, where $c_{x_q} \in \eta \mathbb{Z}^n$ with $\eta \mathbb{Z}^n = \{ c \in \R^n \mid \exists \ell \in \mathbb{Z}^n \text{ s.t. } c = \eta \odot \ell \}$ and $\odot$ denoting element-wise multiplication. The remaining cells in $X_q \setminus \overline{X}_q$ are called overflow cells \cite{reissig2011computing}. 
    \item The initial set is given by $X_{0,q} = X_q \cap X_{0,h}$, and $U_q \subset U_h$ is a finite subset of the input space $U_h$.
    \item Let $A := \left\{ x_q' \in \overline{X}_q \mid x_q' \cap {\mathsf{Reach}}_h(x_q, u_q) \neq \emptyset \right\}$, for $x_q \in \overline{X}_q$ and $u_q \in U_q$. If $A \subseteq \overline{X}_q$, then $Post_u(x_q) := A$; otherwise, $Post_u(x_q) := \emptyset$. Moreoever, $Post_u(x_q) := \emptyset$, for all $x_q \in X_q \setminus \overline{X}_q$.
\end{itemize}
A more detailed discussion can be found in \cite{SCOTS}.

\begin{thm}[{\cite[Theorem VIII.4]{Reissig2017feedback}}]
Let $\Sigma_q(\V)$ be the symbolic model of $\Sigma_h(\V)$ with quantization parameter $\eta \in (\R^+)^n$, Then, the relation $Q := \{ (x_h, x_q) \in X_h \times X_q \mid x_h \in x_q \}$ is a feedback refinement relation from $\Sigma_h(\V)$ to $\Sigma_q(\V)$, i.e., $\Sigma_h(\V) \preceq_Q \Sigma_q(\V)$.
\end{thm}

\subsubsection{Controller Synthesis and Refinement}
Given the sampled system with sampling time $h$, we define the modified specification $\phi_{\lambda+\delta}$ as a sequence of RAS tasks,  
$\phi_{\lambda+\delta} = \phi^1_{\lambda+\delta} \rightarrow \phi^2_{\lambda+\delta} \rightarrow \ldots \rightarrow \phi^{N_s}_{\lambda+\delta},$
where each sub-task is expressed as  
$\phi_{\lambda+\delta}^i := \eventually G_{\lambda+\delta}^i \wedge \always (X_{\lambda+\delta} \setminus O_{\lambda+\delta}^i), \quad i \in \{1,\dots,N_s\}.$
Here, $\delta = \frac{\bar{u}h}{2} + \eta$ denotes the robustness margin, where $\eta$ is the quantization parameter and $\bar{u}$ is the input bound. This extra $\delta$ margin reflects the idea of contracting and expanding atomic propositions as used in \cite{fainekos2009temporal}, and is needed to account for the continuous-time trajectory, including inter-sample intervals, as noted in \cite[Example 2]{liu2014abstraction}.

The corresponding abstract specification $\hat{\phi}_{\lambda+\delta}$ is obtained by translating each reach-avoid-stay sub-task into the abstract domain:  
$$\hat{\phi}_{\lambda+\delta}^i := \eventually \hat{G}_{\lambda+\delta}^i \wedge \always  (\hat{X}_{\lambda+\delta} \setminus \hat{O}_{\lambda+\delta}^i), \quad i \in \{1,\dots,N_s\},$$
where the abstract sets are defined as  
\begin{align*}
    \hat{G}^i_{\lambda+\delta} &:= \{ z_q \in \overline{X}_q \mid Q^{-1}(z_q) \subseteq G^i_{\lambda+\delta} \}, \\    
    \hat{O}^i_{\lambda+\delta} &:= \{ z_q \in \overline{X}_q \mid Q^{-1}(z_q) \subseteq O^i_{\lambda+\delta} \}, \\
    \hat{X}_{\lambda+\delta} &:= \{ z_q \in \overline{X}_q \mid Q^{-1}(z_q) \subseteq X_{\lambda+\delta} \}.
\end{align*}

Now, we address the problem of designing a controller $C_h$ for $\Sigma_h(\V)$ that satisfies the modified specification $\phi_{\lambda+\delta}$. 

\begin{defn}[Specification satisfaction]

Let $P$ be a finite set of atomic propositions that label the state space $X \subset \R^n$ through a labeling function $L: X \rightarrow 2^P$. Consider an LTL specification $\phi_{\lambda+\delta}$ over $P$.

For the transition system $\Sigma_h(\V)$, the finite or infinite run starting from an initial state $x_0 \in X_{0,h}$ is written as
$$x^{(h)}_{x_0,u} := x_0 \xrightarrow[h]{u_0} x^{(h)}_1 \xrightarrow[h]{u_1} x^{(h)}_2 \xrightarrow[h]{u_2} \ldots$$ 
where each transition $x^{(h)}_i \xrightarrow[h]{u_i} x^{(h)}_{i+1}$ with $x^{(h)}_i \in X_h \subseteq X$ for all $i \in \N_0$ and $u = u_0, u_1, u_2, \ldots$.
The system is said to satisfy the specification $\phi_{\lambda+\delta}$, denoted $\Sigma_h(\V) \mid C_h \models \phi_{\lambda+\delta}$, if the controlled trace $L(x_{x_0,u}):=L(x_0),L(x_1),\ldots $, such that $x_{i+1}\in Post_{u_{C,h}}(x_{i}) $ with $u_{C,h}\in C_h(x_i)$, satisfies $ \phi_{\lambda+\delta}$, i.e., $L(x_{x_0,u}) \models \phi_{\lambda+\delta}$. A formal definition of specification satisfaction can be found in \cite[Definition VI.1]{Reissig2017feedback}.
\end{defn}

Because of the feedback refinement relation between $\Sigma_h(\V)$ and its symbolic model $\Sigma_q(\V)$, we can therefore reduce the problem to the finite abstraction $\Sigma_q(\V)$ with respect to the abstract specification $\hat{\phi}_{\lambda+\delta}$. Let $C_q$ be the controller for $\Sigma_q(\V)$ and $\hat{\phi}_{\lambda+\delta}$. Since $\Sigma_q(\V)$ has finitely many states and inputs, $C_q$ can be computed using the standard fixed-point algorithms \cite{tabuada2009verification}. Once $C_q$ is obtained, it can be directly refined into a controller for $\Sigma_h(\V)$ that ensures satisfaction of $\phi_{\lambda+\delta}$. Furthermore, following the results in \cite{liu2016finite}, \cite{liu2014abstraction}, this guarantee extends to the continuous-time system, ensuring that $\Sigma(\V) \models \phi_\lambda$, as stated in the following theorem.

\begin{thm}[Symbolic Controller Refinement]\label{thm:symbol}
Let $\Sigma_h(\V) \preceq_Q \Sigma_q(\V)$ and let $C_q$ be a symbolic controller such that
$\Sigma_q(\V) \mid C_q \models \hat{\phi}_{\lambda+\delta}$. Then, the refined controller
$C_h := C_q \circ Q$ guarantees that the sampled system satisfies the modified specification, i.e.,
$\Sigma_h(\V) \mid C_h \models \phi_{\lambda+\delta}$. Furthermore, this ensures that the continuous-time system satisfies the original modified specification, i.e., $\Sigma(\V) \mid C_h \models \phi_\lambda$.
\end{thm}

\begin{pf}
See Appendix \ref{sec:symbol_proof}
\end{pf}
Existing toolboxes such as \cite{Mazo2010}, \cite{SCOTS}, \cite{khaled2021omegathreads}, \cite{jagtap2017quest} are available for such symbolic controller synthesis.

\begin{rem}
   Traditional symbolic control does not guarantee correctness between sampling instants. The approaches in \cite{liu2016finite}, \cite{liu2014abstraction} address this issue by contracting and expanding atomic propositions with a robustness margin $\delta$. However, this margin is directly proportional to the Lipschitz constant of the system, which can make the method conservative.   In our case, the VCZ follows a simple single-integrator dynamics \eqref{eqn:vcz}. This allows us to compute $\delta$ directly using the control bound $\bar{u}$, which greatly reduces conservatism. Importantly, this does not require any knowledge of the system dynamics, unlike the approach in \cite{liu2016finite}. The conservatism in $\delta$ can be reduced even further by using smaller sampling periods $h$ and finer quantization $\eta$.

   It is also important to note that although the symbolic approach struggles with high-dimensional nonlinear systems, the simplicity of the single-integrator VCZ dynamics allows efficient computation of reachable sets and transitions during abstraction.
\end{rem}


\begin{contexample}{Running Example}
    For the VCZ $\mathcal{B}(\xi(t), \lambda)$ corresponding to the pendulum system, we synthesize a symbolic controller using the SCOTS toolbox \cite{SCOTS} within the input bounds $\bar{u} = 0.1$~rad/s. The controller enforces the modified specification $\phi_\lambda = \xi(t) \in [-0.182, 0.182]$~rad, for all $t \in \R_0^+$. Figure~\ref{fig:vcz_pend3} shows the resulting VCZ trajectory $\xi(t)$ over time, which remains within the specified bounds, confirming that $\phi_\lambda$ is satisfied. \qed
\end{contexample}

\section{Control for Confinement Within the VCZ} \label{Sec:Contorl_for_VCZ}
\RD{Unlike standard abstraction-based refinement, the symbolic controller is not directly applied to the original system. Instead, it is refined through a closed-form confinement controller that enforces the virtual behavior on the original dynamics under bounded uncertainty.}

Given the EL system in Equation~\eqref{eqn:sysdyn} and its corresponding VCZ $\mathcal{B}(\xi(t), \lambda)$, we design a closed-form, model-free control input $\tau(t)$. We further derive feasibility conditions that account for bounds on system parameters and input constraints. These conditions guarantee that the system configuration $x(t)$ remains within the VCZ under the applied control input, i.e.,
\begin{equation}\label{eqn:vcz_bounds}
    \|x(t) - \xi(t)\| < \lambda, \quad \forall t \in \R_0^+.
\end{equation}
Finally, we analyze the trade-off between control efficiency and conservatism in the choice of the VCZ radius $\lambda$ and the virtual input bound $\bar{u}$.

\subsection{Controller Design}
The proposed controller is derived using a methodical two-step approach, a velocity-level control \eqref{eqn:sysDyn_vel} to ensure confinement, and then an acceleration-level control \eqref{eqn:sysDyn_acc} to track the desired reference velocity. The EL system in \eqref{eqn:sysdyn} can be re-written in state-space representation as:
\begin{subequations}
    \begin{align}
    \dot{x} &= v, \label{eqn:sysDyn_vel}   \\
    \dot{v} 
    &= V_M(x,v) + M(x)^{-1}\tau + M(x)^{-1}d, \label{eqn:sysDyn_acc}
    \end{align}
\end{subequations}
where $V_M(x,v) = -M(x)^{-1}(V(x,v)+G(x))$.
The two-step backstepping-like approach draws inspiration from the model-free funnel control framework discussed in \cite{PPCfeedback} and \cite{hard_soft}, where it was used to handle pure feedback systems and EL systems, respectively.

\subsubsection{Stage I}
Given the VCZ $\B (\xi(t),\lambda)$, centered at $\xi(t)$ with radius $\lambda\in \R^+$, $e_x(t) := (\|x(t) - \xi(t)\|) / \lambda$ is the normalized position error.
To keep $x(t)$ confined within the VCZ, we define the velocity-level control input $v_r(t)$ as 
\begin{equation} \label{eqn:velcon}
    v_r(t) = -\overline{v} \Psi(e_x) \mathbb{I}_{n} \frac{x(t) - \xi(t)}{\|x(t) - \xi(t)\|},
\end{equation}
where $\overline{v} \in \R^n$ is the maximum allowed velocity and $\Psi:\R \rightarrow \R$ is a transformation function (Appendix \ref{sec:clamp}).

This velocity control law ensures that the system configuration $x$ is \textit{pulled} toward the center $\xi(t)$ while always staying within the VCZ (Equation~\eqref{eqn:vcz_bounds}). 

\subsubsection{Stage II}
To make sure the actual system velocity $v(t)$ tracks the reference velocity $v_r(t)$ in Equation in \eqref{eqn:velcon}, we define the velocity tracking error as:
$e_v(t) = v(t) - v_{r}(t).$ 
We enforce that $e_v(t)$ remains bounded within exponentially decaying funnel constraints $\rho_v: \R_0^+ \rightarrow \R^n$, given by:
\begin{equation}\label{eqn:funnel_v}
\rho_v(t) = e^{-\mu_v t}(p_v-q_v) + q_v, 
\end{equation}
where $p_v \in \R^n$ defines the initial width of the funnel (with $|e_v(0)| \preceq p_v$), $q_v \in \R^n$ defines the steady-state bound (with $0 \prec q_v \prec p_v$), and $\mu_v \in \R^{n \times n}$ is a diagonal matrix that controls how quickly the funnel contracts.

The velocity error $e_v$ is constrained within the funnel: 
\begin{equation}\label{eqn:fun2}
    -\rho_v(t) \prec e_v(t) \prec \rho_v(t).    
\end{equation}
The control input $\tau(t)$ at the acceleration level is:
\begin{equation}\label{eqn:con}
    \tau(t) = -\overline{\tau}\text{diag}(\Psi(\varepsilon_v)),
\end{equation}
where $\overline{\tau} \in \R^n$ is the maximum allowable torque, and $\Psi(\cdot)$ is the transformation function (Appendix \ref{sec:clamp}) and $\varepsilon_v(t) := \text{diag}(\rho_{v})^{-1}e_v(t)$ is the normalized velocity error.



\begin{figure*}[t]
    \centering
    \includegraphics[width=0.95\textwidth]{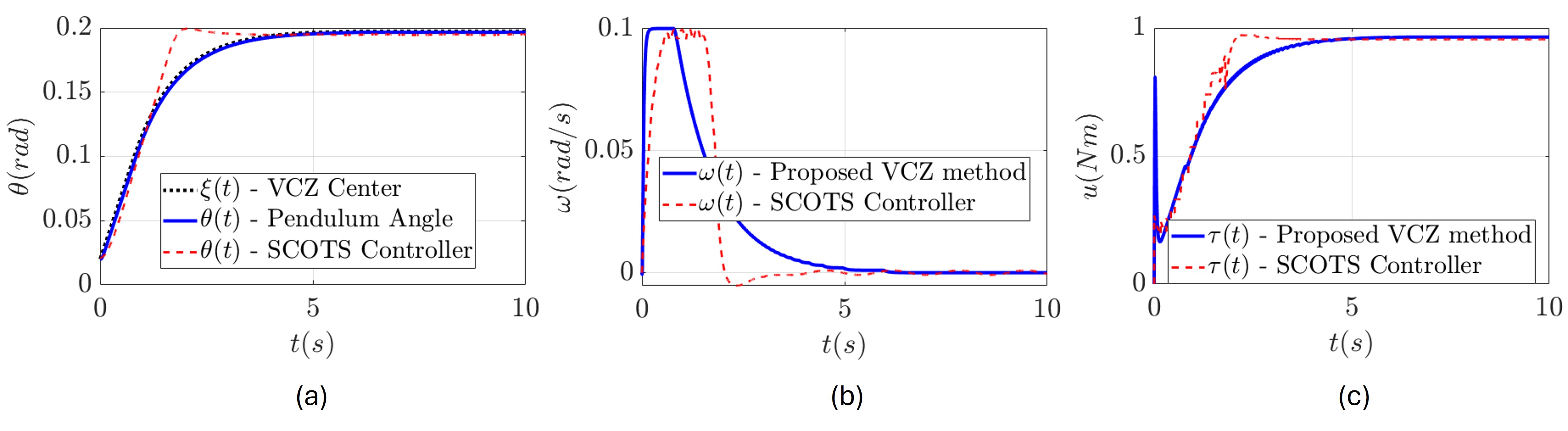}
      \caption{Pendulum system: (a) VCZ center $\xi(t)$ and pendulum angle $\theta(t)$; (b) angular velocity $\omega(t)$ and VCZ control $u(t)$; and (c) control torque $\tau(t)$, under the proposed VCZ control and symbolic control.}
    \label{fig:vcz_pend3}
\end{figure*}

\begin{contexample}{Running Example}
Revisiting the pendulum example, we rewrite the dynamics as:
\begin{align}\label{eqn:pend2}
    \dot{x} &= v, \\
    \dot{v} 
            &= -\frac{3g}{2l}\sin x + \frac{3}{ml^2}\tau + \frac{3}{ml^2}d.
\end{align}
To enforce $x(t) \in \mathcal{B}(\xi(t), \lambda)$ with $\xi(t) \in \R$ and $\lambda = 0.018$ rad, the reference velocity is generated via Equation~\eqref{eqn:velcon} with $\bar{v} = 0.1$~rad/s. The control input $\tau$ is then computed from Equation~\eqref{eqn:con}, with the bound $\bar{\tau} = 5$~Nm. Figure~\ref{fig:vcz_pend3} shows the time evolution of the VCZ trajectory $\xi(t)$ and its control signal $u(t)$, together with the pendulum’s states $[\theta(t), \omega(t)]^\top$ and the applied control torque $\tau(t)$. \qed
\end{contexample}

\subsection{Feasibility Conditions}\label{sec:feas}
\RD{To ensure confinement within the VCZ, we derive feasibility conditions that account for system parameter uncertainties and input constraints. 
This step connects the symbolic controller synthesized for the virtual VCZ dynamics to the original EL system. 
Although we synthesize the controller for a simplified virtual system, the feasibility conditions in this section ensure that the real continuous-time system can follow the VCZ trajectory despite bounded disturbances and input constraints.}

Given the EL dynamics in \eqref{eqn:sysdyn}, under Assumptions~\ref{assum:unknown dynamics} and \ref{assum:sys_bounds}, the VCZ $\B (\xi(t),\lambda)$, the funnel constraint \eqref{eqn:funnel_v}, and the input bound $\bar{u}$, the maximum permissible velocity $\overline{v}$ and torque $\overline{\tau}$ should adhere to the following conditions:
\begin{subequations}\label{eqn:feas}
\begin{align}
    \overline{v} &\succeq \overline{u}, \label{eqn:feas1} \\
    \overline{\tau} &\succeq \frac{1}{\underline{m}} \left( V_M^{\max} + \underline{m}_i\overline{d} + \mu_v(p_v-q_v) + \overline{a}_r \right), \label{eqn:feas2}
\end{align}
\end{subequations}
where using the bounded transformation function provided in Appendix \ref{sec:clamp}, we compute an upper bound $\overline{a}_r \in \R^n$, such that $|\dot{v}_r| \preceq \overline{a}_r$, as detailed in Appendix \ref{sec:bd_vr}.

\begin{rem}
\RD{ The feasibility conditions in \eqref{eqn:feas} reflect the fundamental requirement that the available actuation must be sufficient to keep the system confined within the VCZ while compensating for static system effects (gravity and Coriolis terms) and bounded external disturbances. }
\end{rem}

\begin{contexample}{Running Example}
For the pendulum system, with the choice of input bounds set at $\bar{v} = \bar{u} = 0.1$~rad/s, the first feasibility condition in \eqref{eqn:feas1} is satisfied.\\
To verify the second feasibility condition \eqref{eqn:feas2}, we define the transformation function $\Psi(s) := \tanh^3(as)$ with $a = 1.8$, and with the VCZ radius set to $\lambda = 0.018$~rad, we obtain $a_r = 2.25\bar{v}(\bar{v}+\bar{u})/\lambda = 2.5$~rad/s$^2$ (Appendix~\ref{sec:bd_vr}). \\
For the funnel constraint in \eqref{eqn:fun2}, we select the parameters $p_v = 1$, $q_v = 0.01$, and $\mu_v = 1$. Plugging these into the right-hand side of the condition gives $\frac{1}{3} \left( 1 + 1.5 + 0.99 + 2.5 \right) = 1.997,$ showing that the second condition holds, as $\bar{\tau} = 2$~Nm.   \qed 
\end{contexample}

\subsection{Efficiency vs Conservativeness}\label{sec:lam-ub}
The two key parameters introduced for symbolic control of the VCZ are the radius $\lambda$, discussed in Section~\ref{sec:vcz}, and the control input bound $\bar{u}$, introduced in Section~\ref{sec:symbolic}. Now, we examine the trade-off between these two parameters.

Note that in the feasibility conditions, while the system parameter bounds and the input bounds, $\bar{v}, \bar{\tau}$ are given, we can choose the funnel parameters $p_v, q_v, \mu_v$, the VCZ radius $\lambda$, and the maximum velocity of the VCZ $\bar{u}$ to not only meet the feasibility conditions, but also maximize the performance.

For $\Psi(s) = \tanh^3(as)$ with $a = 1.8$, the corresponding parameter $a_r$ is given by $a_r = \frac{2.25\bar{v}(\bar{v}+\bar{u})}{\lambda}$ (Appendix~\ref{sec:bd_vr}). Therefore, we can rewrite \eqref{eqn:feas2} as,
\begin{align*}
    \bar{u} \preceq \frac{\lambda}{2.25 \bar{v}} \Big( \underline{m}\bar{\tau} -  V_M^{\max} - \underline{m}_i\overline{d} - \mu_v(p_v-q_v)\Big) - \bar{v}.    
\end{align*}
Combining with first feasibility condition \eqref{eqn:feas1}, we get
\begin{align}\label{eqn:lam_ubar}
    \bar{u} \preceq \min \Big( \bar{v}, \frac{\lambda}{2.25 \bar{v}} ( \underline{m}\bar{\tau} -  V_M^{\max} - \underline{m}_i\overline{d} - \mu_v(p_v-q_v) ) - \bar{v} \Big).    
\end{align}
We now identify two viable strategies for selecting the VCZ parameters, depending on the desired trade-off between conservatism and control efficiency:

\textbf{A. Least conservative VCZ:} 
A larger VCZ radius $\lambda$ allows the system more room to deviate from the center, requiring a more conservative modification of the original specification. This typically means inflating obstacle sizes and shrinking target sets (Equation~\eqref{eqn:redspec}), leading to overly cautious behavior. To minimize this conservatism, the smallest feasible $\lambda$ is obtained by solving:
\begin{align}\label{eqn:least_conservative}
    \frac{\lambda}{2.25 \bar{v}} \Big( \underline{m}\bar{\tau} \!-\!  V_M^{\max} \!-\! 
    \underline{m}_i\overline{d} \!-\! \mu_v(p_v \!-\! q_v) \Big) - \bar{v} \!\succ\! 0.    
\end{align}
\RD{However, minimizing $\lambda$ also decreases the input bound, $\bar{u}$, resulting in slower and more cautious task execution.}

\textbf{B. Most efficient VCZ:}
The objective is to minimize $\lambda$ while maximizing $\bar{u}$ to achieve fast and effective control without excessive conservatism. If $\bar{u}$ is chosen too small, the VCZ will be slow regardless of the system’s actual capabilities. To make full use of the available input range, we set $\bar{u} = \bar{v}$, and solve:
\begin{align}\label{eqn:most_efficient}
    \bar{v} \!=\! \frac{\lambda}{2.25 \bar{v}} ( \underline{m}\bar{\tau} \!-\!  V_M^{\max} 
    \!-\! \underline{m}_i\overline{d} \!-\! \mu_v(p_v-q_v)) \!-\! \bar{v}.    
\end{align}

\begin{contexample}{Running Example}
To better understand this trade-off, we revisit the pendulum example. 
From Equation~\eqref{eqn:lam_ubar}, we obtain:
\begin{align*}
    \bar{u} &\leq \min(0.1, \frac{\lambda}{0.225}(3(2) - 1 - 1.5 - 0.99) - 0.1) \leq \min(0.1, 11.156\lambda-0.1).
\end{align*}
\textbf{A. Least conservative VCZ:} To ensure $\bar{u} > 0 \implies 11.156\lambda-0.1 > 0$, we must choose $\lambda > 0.009$~rad. If we select 
$\lambda = 0.01\text{ rad, then } \bar{u} \leq 0.01 \text{ rad/s.}$ 
Although the pendulum can move at up to $\bar{v} = 0.1$~rad/s, the VCZ is limited to move at one-tenth that speed, causing the system to operate significantly below its capability. 

\textbf{B. Most efficient VCZ:} To match the VCZ with the system’s maximum velocity $\bar{u} = \bar{v} = 0.1$~rad/s, we get:
$$0.1 = 11.156\lambda-0.1 \implies \lambda = 0.018 \text{ rad}, \bar{u} = 0.1 \text{ rad/s.}$$ 
By slightly increasing the VCZ radius, we allow it to fully leverage the system's velocity bound, enabling faster task completion with minimal additional conservatism. \qed   
\end{contexample}

In summary, we can choose any combination of $\lambda$ and $\bar{u}$ within this feasible region to suit our control objectives, whether it is prioritizing tighter safety margins or faster task execution. This flexibility is a key strength of the proposed framework.

The following theorem formally summarizes the control law $\tau(t)$ proposed in this paper.

\begin{thm}\label{thm:mainresult}
    Consider the EL system $\mathcal{S}$ in \eqref{eqn:sysdyn}, subject to Assumptions~\ref{assum:unknown dynamics} and \ref{assum:sys_bounds}, and assigned a temporal logic specification $\phi$. Suppose the initial error satisfies $\|e_x(0)\| < \lambda$, and the feasibility conditions \eqref{eqn:feas} hold. 
    Then the closed-form confinement controller
    \begin{align}\label{eqn:control}
        \tau(t) &= -\overline{\tau} \Psi \Big( \text{diag}(\rho_v)^{-1} \Big( \dot{x}(t) + \overline{v} \Psi \Big( \frac{x(t) - \xi(t)}{\|x(t) - \xi(t)\|} \Big)
        \mathbb{I}_n \frac{x(t) - \xi(t)}{\|x(t) - \xi(t)\|} \Big) \Big)
    \end{align}
    guarantees that the system configuration remains strictly within the VCZ for all time, i.e., 
    $$\|x(t) - \xi(t)\| < \lambda, \forall t \geq 0.$$
    Here, $\xi(t)$ is the VCZ centre, which evolves according to 
    $$\dot{\xi}(t) = C_q(Q(\xi(t))),$$
    where $Q$ is the feedback refinement relation, and $C_q$ is the symbolic controller from Theorem~\ref{thm:symbol}, ensuring  $\Sigma(\mathcal{V}) \mid (C_q \circ Q) \models \phi_\lambda$.
    
    Consequently, by Theorem~\ref{thm:red}, satisfaction of the modified specification $\phi_\lambda$ by the VCZ dynamics, together with confinement of $x(t)$ within the VCZ, implies satisfaction of the original specification $\phi$ by $\mathcal{S}$.
\end{thm}
\begin{pf}
    See Appendix \ref{sec:cont_proof} for the proof.
\end{pf}

\begin{contexample}{Running Example}
Therefore, under the control law in Equation~\eqref{eqn:con}, 
$x(t)$ remains inside the VCZ, i.e., $\|x(t)-\xi(t)\| < \lambda$ or $x(t) \in \B(\xi(t), \lambda)$ for all $t \in \R_0^+$. The symbolic controller from Section~\ref{sec:symbolic} ensures that the center trajectory $\xi$ satisfies the modified specification $\xi(t) \in [-0.182, 0.182]$~rad for all $t \in \R_0^+$. By Theorem~\ref{thm:red}, this guarantees that the actual system $x(t)$ meets the original specification, i.e., $x(t) \in [-0.2, 0.2]$ for all $t \in \R_0^+$, as shown in Figure~\ref{fig:vcz_pend3}. \qed   
\end{contexample}

\begin{rem} 
The proposed framework offers several advantages over traditional abstraction-based control. 
\RD{First, the adoption of simple single-integrator VCZ dynamics in Equation~\eqref{eqn:vcz} ensures that the abstraction and controller synthesis in Theorem~\ref{thm:symbol} remain \textit{scalable}.}
Second, Theorem~\ref{thm:symbol} shows that the robustness margin $\delta$ provides \textit{inter-sample guarantees} for continuous-time trajectories. 
\RD{Finally, the closed-form confinement controller in \eqref{eqn:control}, established in Theorem~\ref{thm:mainresult}, is \textit{approximation-free} in the sense that it is model-free with respect to EL parameters. While the system is assumed to have an EL structure, the controller does not require identifying, learning, or approximating the matrices $M,V,G$. Instead, it relies only on known bounds on parameters, disturbances and input limits to provide formal guarantees.}
Together, these properties allow continuous-time specification satisfaction under model uncertainties and bounded disturbances, while significantly reducing computational complexity. These advantages are further highlighted through case studies in the next section.
\end{rem}

\section{Case Studies}\label{Sec:Case Study}

To demonstrate the effectiveness and scalability of the proposed approach, we consider two additional case studies. The first involves a 4-dimensional two-link SCARA manipulator, while the second focuses on an 8-dimensional multi-agent system, highlighting the method’s capability to higher-dimensional scenarios. We also present a comparative study against traditional symbolic control solutions implemented using the SCOTS toolbox \cite{SCOTS}.
\RD{All computations and experiments were performed on Ubuntu~20.04 running on a system with an Intel Core i9-10900 processor and 64~GB RAM. All the codes are available at \href{https://github.com/FocasLab/VCZ.git}{https://github.com/FocasLab/VCZ.git}.}

\begin{figure*}[t]
    \centering
    \includegraphics[width=\textwidth]{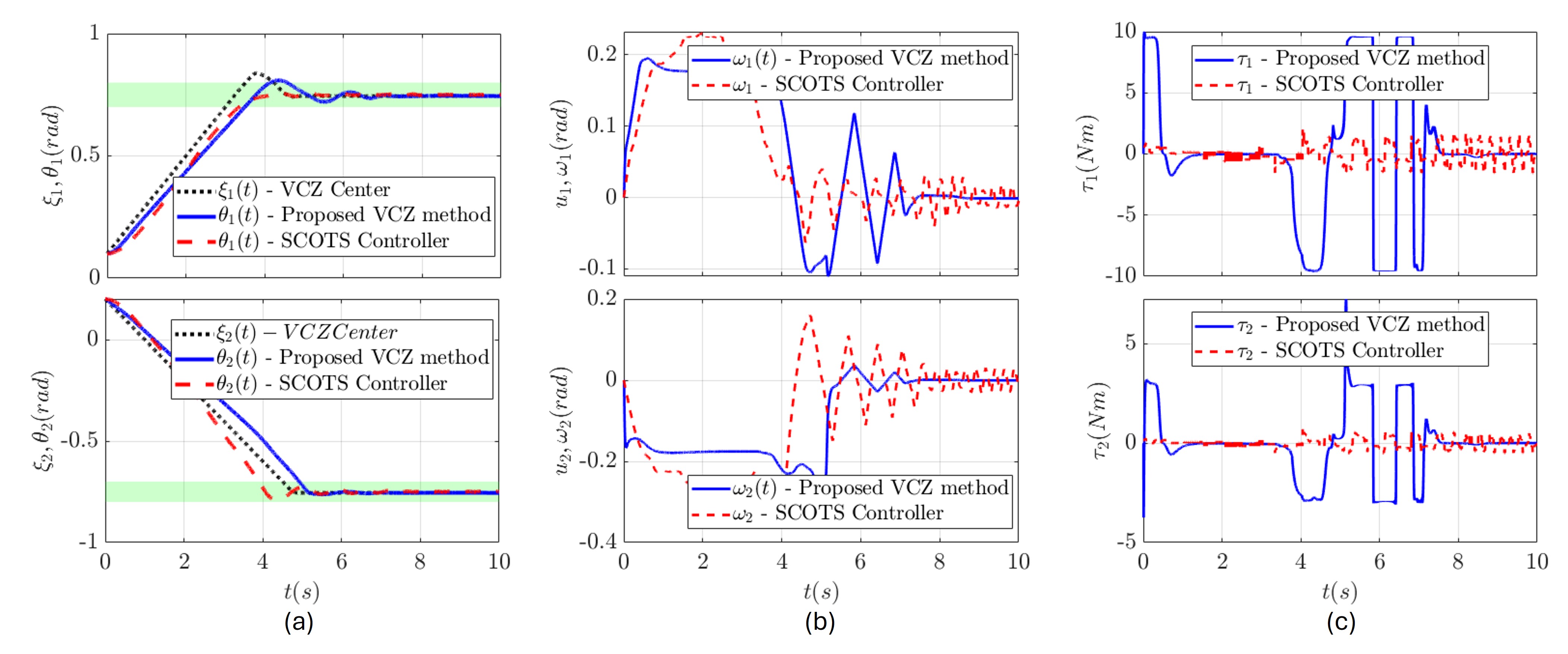}
      \caption{Two-link SCARA manipulator system: (a),(d) VCZ center $\xi(t)$ and joint angles $\theta(t)$; (b),(e) angular velocities $\omega(t)$; and (c),(f) control torques $\tau(t)$, under the proposed VCZ control and symbolic control.}
    \label{fig:vcz_2r}
\end{figure*}

\subsection{Two-link SCARA manipulator}
We now consider a 4-dimensional two-link planar SCARA manipulator \cite{craig2009introduction} with two revolute joints. The system configuration is given by the joint angles $x(t) = [\theta_1(t), \theta_2(t)]^\top$ and its dynamics are described by
\begin{gather*}
    ml^2
    \begin{bmatrix}
        \frac{5}{3} + c_2 & \frac{1}{3} + \frac{1}{2}c_2 \\
        \frac{1}{2}c_2 & \frac{1}{3}
    \end{bmatrix}
    \begin{bmatrix}
        \Ddot{\theta}_1 \\
        \Ddot{\theta}_2
    \end{bmatrix} 
    +
    ml^2s_2
    \begin{bmatrix}
        -\frac{1}{2}\dot{\theta}_2^2 - \dot{\theta}_1\dot{\theta}_2\\
        \frac{1}{2}\dot{\theta}_2^2
    \end{bmatrix}
    + mgl
    \begin{bmatrix}
        \frac{3}{2}c_1 + \frac{1}{2}c_{12} \\
        \frac{1}{2}c_{12}
    \end{bmatrix}
    = 
    \begin{bmatrix}
        \tau_1(t) \\
        \tau_2(t)
    \end{bmatrix}
    + d(t), \nonumber
\end{gather*}
where \RD{$m$ and $l$ are the unknown link mass and length}, $g$ is the acceleration due to gravity, \RD{$d(t)$ is an unknown disturbance}, and $\tau_1(t), \tau_2(t)$ are the joint torque inputs. $c_1 := \cos{\theta_1}$, $c_2 := \cos{\theta_2}$, $s_2 := \sin{\theta_2}$, and $c_{12} := \cos{(\theta_1 + \theta_2)}$. 

The task specification for the manipulator is given by 
$\phi = \eventually G$,
where the target set is $G = [0.7,0.8] \times [-0.8, -0.7]$. This specification requires that the joint configuration eventually reaches the region $G$.

The physical parameters are assumed unknown, but the following bounds from Assumption~\ref{assum:sys_bounds} hold: $\underline{m}=1.5 \text{ kg}, \underline{m}_i = 1.6 \text{ kg}^{-1}$ and $\max{(-\underline{V}_M, \overline{V}_M)} = 5 m^2/s^2$. The disturbance is bounded by $\overline{d} = 0.2$~Nm. The maximum joint velocity is $\overline{v} = 0.2$~rad/s and maximum torque is $\overline{\tau} = 10$~Nm.
For the acceleration-level funnel, we choose parameters $p_x = 0.1, q_x = 0.01,$ and $\mu_x = 0.1$. The VCZ radius is set to $\lambda = 0.019$~rad as per \eqref{eqn:most_efficient}. These values satisfy the feasibility conditions \eqref{eqn:feas}.

The VCZ trajectory is obtained via symbolic control, and Figure~\ref{fig:vcz_2r} shows the resulting system trajectory, VCZ center, and control effort.

\begin{figure*}[t]
    \centering
    \includegraphics[width=0.85\textwidth]{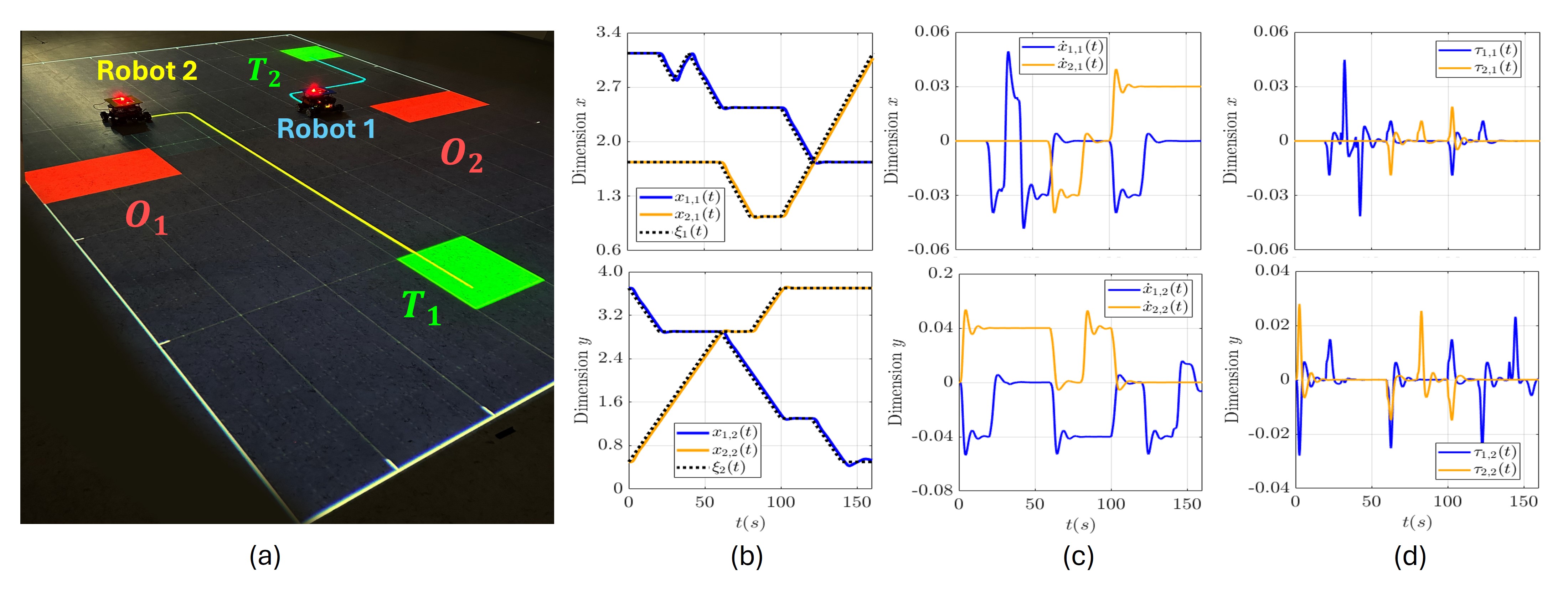}
      \caption{Multi Agent system: (a) System trajectory; (b),(d) VCZ center $\xi(t)$ and system configuration $x(t)$; (c),(e) control input $\tau(t)$, under the proposed VCZ control and symbolic control.
      \href{https://indianinstituteofscience-my.sharepoint.com/:f:/g/personal/ratnangshud_iisc_ac_in/Eh4Nf7uP7XNOv8P0j9iY4VcBPufmRohNqDteC-UqboETqQ?e=UaDbOa}{[Video]}}
    \label{fig:vcz_multi}
\end{figure*}

\subsection{Multi-Agent Systems}
In the third case study, we consider an 8-dimensional multi-agent setup with two robots operating in a two-dimensional environment. Each robot $i \in \{1,2\}$ is modeled by the second-order dynamics
\begin{equation}
[\Ddot{x}_1 \ \Ddot{x}_2]^\top = [\tau_1(t) \ \tau_2(t)]^\top + d(t),
\end{equation}
where $x_i = [x_{i,1}, x_{i,2}]^\top \in \R^2$ denotes the position of agent $i$, $\tau_i(t) = [\tau_{i,1}, \tau_{i,2}]^\top \in \R^2$, $i \in \{1,2\}$ is its acceleration input, and \RD{$d(t) \in \R^4$ is an unknown disturbance}.

\RD{The specification for the multi-agent setup simultaneously enforces goal-reaching, persistent obstacle avoidance, and continuous inter-agent safety for both agents. It is defined as $\phi = \phi_1 \wedge \phi_2$, where the specification for each agent $i \in \{1,2\}$ is given by}
$\phi_i = \eventually T_i \wedge \always \neg (O_1 \wedge O_2) \wedge \always D$. 
Here $T_i$ denotes the goal region of agent $i$, $O_1$ and $O_2$ represent two static obstacles in the environment, $D$ enforces a minimum inter-agent separation, and the temporal operators $\always$ and $\eventually$ correspond to “always” and “eventually,” respectively.

Using the VCZ-based symbolic control framework, we synthesize controllers for both agents to satisfy the LTL specification $\phi$. Each agent is constrained by a maximum velocity of $\overline{v} = 0.2$~m/s and a maximum acceleration of $\overline{\tau} = 0.2$~m/s$^2$. The acceleration-level funnel parameters are set to $p_x = 0.1$, $q_x = 0.01$, and $\mu_x = 0.1$, with the VCZ radius chosen as $\lambda = 0.8$~m.

The target regions $T_i$, the obstacles $O_1$ and $O_2$, and the resulting system trajectories are illustrated in Figure 3. The figure also shows the time evolution of the VCZ centers, the closed-loop trajectories, and the corresponding control inputs. The results demonstrate that both robots successfully reach their respective targets without colliding with the obstacles or each other, thereby satisfying the LTL specification.

\begin{table*}[t]
\centering
\caption{Comparison of computation time, memory, and controller domain size between the VCZ approach and the symbolic approach.}
\begin{tabularx}{\textwidth}{l *{8}{>{\centering\arraybackslash}X}}
\hline
\multirow{2}{*}{\textbf{Case Study \qquad\qquad\qquad}} 
& \multicolumn{3}{c}{\textbf{Computation Time (s)}} 
& \multicolumn{3}{c}{\textbf{Memory (kB)}} 
& \multicolumn{2}{c}{\textbf{Domain Size}} \\ \cline{2-9}
& SCOTS & VCZ & \% Red. 
& SCOTS & VCZ & \% Red.
& SCOTS & VCZ \\ \hline
Pendulum (2 dimensions) & 2.1993 & \textbf{0.048} & \textbf{97.82}\% 
         & 226.72 & \textbf{5.94}  & \textbf{97.38}\%
         & 3714   & \textbf{3590} \\
Two-link (4 dimensions) & 1394.25 & \textbf{2.16} & \textbf{99.85}\%
         & 26246.85 & \textbf{93.44} & \textbf{99.64}\%
         & 95149  & \textbf{80640} \\
Multi-Agent (8 dimensions) & - & \textbf{88.70} & - 
            & - & \textbf{44348} & -
            & - & \textbf{459756} \\ \hline
\end{tabularx}
\label{tab:comp_time_mem}
\end{table*}

\subsection{Comparison and Discussion}

Table~\ref{tab:comp_time_mem} presents a comparison between the proposed VCZ-based framework and traditional symbolic control implemented using the SCOTS toolbox \cite{SCOTS}. Across all case studies, our method significantly reduces both computation time and memory usage. For the pendulum system, the reduction is already substantial, but as the dimensionality of the system increases, moving from the 4-dimensional SCARA manipulator to the 8-dimensional multi-agent system, the improvements become dramatically more prominent. This trend highlights that while symbolic control methods suffer from the curse of dimensionality, leading to exponential growth in computational burden, the proposed VCZ-based approach scales efficiently, maintaining low computational overhead even for high-dimensional systems. These results confirm that our method not only ensures correctness and safety but also provides a scalable improvement over the traditional symbolic control.

Another important distinction lies in the requirements of mathematical model. Traditional symbolic control requires the knowledge of the system dynamics to build accurate abstractions. In contrast, the proposed VCZ framework does not depend on a precise model. It only requires bounds on system parameters and input constraints, and it can also handle unknown but bounded disturbances. This makes the VCZ method not only more efficient, but also more practical for real systems, where exact dynamics are often unavailable.


\begin{figure}[t]
    \centering
    \includegraphics[width=0.45\textwidth]{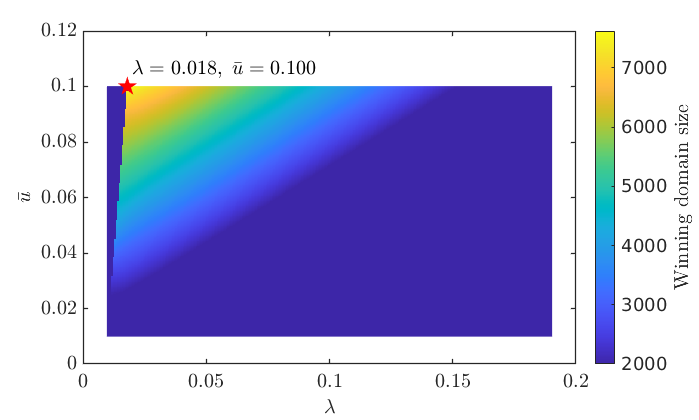}
    \caption{Pendulum system: Winning domain size as a function of the VCZ radius $\lambda$ and admissible velocity $\overline{u}$. The heatmap illustrates the variation in the winning domain, while the red star marks the parameter values that maximize the winning domain.}
    \label{fig:windom}
\end{figure}

\section{Limitations and Future Work}

\RD{The proposed framework also has certain limitations that motivate future research.
First, it assumes fully actuated dynamics and therefore does not directly apply to underactuated or nonholonomic systems, such as aerial vehicles or ground vehicles with nonholonomic constraints. Extending the framework to non-affine systems or more general nonlinear classes of systems is an important direction for future work.
\\
Second, by introducing the parameter $\lambda$ to modify specifications, the VCZ method results in a more conservative solution in terms of the controller’s domain. Consequently, the set of initial conditions from which the task can be guaranteed is smaller than that obtained using SCOTS-based synthesis for the full system, as shown in Table~\ref{tab:comp_time_mem}.
\\
This conservatism is further influenced by the use of a fixed VCZ motion limit across the environment. To reduce this conservatism, future work will explore non-uniform and topology-aware VCZ motion, allowing slower movement in tight regions and faster motion in open areas.
}

\begin{rem}
\RD{Section~\ref{sec:lam-ub} discusses practical choices of the VCZ radius $\lambda$ and the virtual input bound $\bar{u}$, which jointly determine the size of the winning domain. 
Choosing a very small $\bar{u}$ gives the smallest $\lambda$ and the least conservative spatial specification, but it also limits the admissible velocities and results in a smaller winning domain. 
Setting $\bar{u}=\bar{v}$ slightly increases $\lambda$ while allowing a much larger range of velocities, which increases the winning domain. 
Further increasing $\lambda$ without increasing $\bar{u}$ only reduces the admissible state space and shrinks the winning domain. 
The effect of these choices is illustrated for the pendulum example in Fig.~\ref{fig:windom}.}
\end{rem}

\section{Conclusion}
In this work, we introduced a novel symbolic control framework based on Virtual Confinement Zones (VCZs) for enforcing temporal logic specifications in Euler–Lagrange systems. Unlike traditional symbolic control, which requires exact knowledge of the system dynamics, the VCZ approach relies only on bounds on system parameters and input constraints. The proposed approximation-free controller ensures that system trajectories remain within the VCZ, thereby guaranteeing specification satisfaction even in the presence of unknown but bounded disturbances. In addition, the robustness margin in the VCZ radius provides inter-sample guarantees, ensuring correctness for the continuous-time trajectories. Through case studies ranging from low-dimensional systems such as the pendulum to higher-dimensional examples including two-link manipulators and multi-agent systems, we showed that the VCZ framework achieves correctness and safety while significantly reducing computation time and memory usage. Moreover, the results show that the scalability advantage becomes more prominent as the system dimensionality increases, effectively overcoming the curse of dimensionality inherent in symbolic model-based techniques. Overall, the VCZ framework provides a scalable and approximation-free alternative to symbolic control methods, making it well-suited for practical systems where exact models are unavailable and computational efficiency is essential. 

\bibliographystyle{plain}        
\bibliography{autosam}           



\newpage
\appendix{\textbf{\Large Appendix}}
\section{{Bounded Transformation Function}} \label{sec:clamp}
The bounded transformation function $\Psi:\R^n \rightarrow \R^n$ is a smooth mapping that ensures the control inputs remain within specified limits while preserving desired behavior. 
It is defined component-wise $\Psi(s) = [\Psi_1(s_1), \ldots, \Psi_n(s_n)]^\top$, where for all $i = [1;n]$:
$$\Psi_i(s_i) =  
    \begin{cases} 
      -1, & s_i \in (-\infty,-1], \\
      0, & s_i = 0, \\
      1, & s_i \in [1,\infty),
    \end{cases}
$$
and $\Psi_i(s_i)$ is nondecreasing for all $s_i \in (-\infty,\infty)$.
An example of such a saturation transformation function is $\Psi(s) = \tanh^3(as) $ with $a\in\R^+$, and as $a$ increases, we get sharper approximations of the saturation function.

\section{Bounding the Reference Acceleration}\label{sec:bd_vr}
The time derivative of the reference velocity $\overline{v}_r$ \eqref{eqn:velcon} is
\begin{align}
\dot{v}_r(t) = -\overline{v}\Big(\dot{\Psi}(e_x) \hat{e}(t) + \Psi(e_x) \frac{1}{e_x\lambda} \notag \\
\big(\mathbb{I}_n - \hat{e}(t)\hat{e}^\top(t)\big)\big(\dot{x}(t)-\dot{\xi}(t)\big)\Big),
\label{Eq:vel_derivative}
\end{align}
where $e_x=\|x(t)-\xi(t)\|/\lambda$ and $\hat e(t)=(x(t)-\xi(t))/\|x(t)-\xi(t)\|$. 

Time derivative of $\Psi(e_x)=\tanh^3(ae_x)$ (elementwise) is
\begin{align*}
    \dot\Psi(e_x) &=  \frac{d\Psi}{de_x} \frac{1}{\lambda} \frac{(x(t)-\xi(t))^\top(\dot x(t)-\dot\xi(t))}{\|x(t)-\xi(t)\|} \\
    &\le \frac{d\Psi}{de_x}\frac{1}{\lambda} \|\dot x(t)-\dot\xi(t)\|. 
\end{align*}
Solving $\frac{d^2\Psi}{de_x^2}=0$ gives the maximum slope of $\Psi$ (on $e_x\in[0,1]$), and similarly we can bound $\Psi(e_x)/e_x$ on that interval. Numerical values depend on $a$ and for $a = 1.8$,
$\max_{e_x\in[0,1]} \dot \Psi(e_x) \approx 1.35,$ and $\max_{e_x\in(0,1]}\frac{\Psi(e_x)}{e_x} \approx 0.9.$

Using the velocity bound $\|\dot x(t)-\dot\xi(t)\|\le\overline v+\overline u$, we obtain the uniform bound
$$\|\dot v_r(t)\|
\le \overline v \left(\frac{1.35}{\lambda} + \frac{0.9}{\lambda}\right)(\overline v+\overline u) < 2.25 \cdot \overline{v} \cdot \frac{\overline{v} + \overline{u}}{\lambda} =:\overline a_r $$

\section{Proof of Theorem \ref{thm:red}}\label{sec:red_proof}
\begin{pf}
Suppose the VCZ center trajectory $\xi:\R_0^+ \rightarrow X \subset \R^n$ satisfies the modified specification $\phi_\lambda$, and the system configuration $x(t)$ remains confined within the VCZ, i.e., $\|x(t) - \xi(t)\| < \lambda$ for all $t \in \R_0^+$. We aim to show that the system trajectory $x:\R_0^+ \rightarrow X \subset \R^n$ satisfies $x \models \phi$.

Recall that the LTL specification $\phi$ can be decomposed into an ordered sequence of reach-avoid-stay subtasks $\phi = \phi^1 \rightarrow \phi^2 \rightarrow \ldots \rightarrow \phi^{N_s}$, where each $\phi^i$ is defined as:
$
\phi^i := \eventually G^i \wedge \always (X \setminus O^i),
$
and the modified specification is given by $\phi_\lambda = \phi^1_\lambda \rightarrow \phi^2_\lambda \rightarrow \ldots \rightarrow \phi^{N_s}_\lambda$, where each $\phi^i_\lambda$ is:
$
\phi_\lambda^i := \eventually G_\lambda^i \wedge \always (X_\lambda \setminus O_\lambda^i),
$
with:
\begin{align*}
    G_\lambda^i &:= \{ z \in G^i \mid z' \in G^i, \forall z' \text{ with } \|z - z'\| \leq \lambda \}, \\   
    O_\lambda^i &:= \{ z \in \R^n \mid \|z - z'\| \leq \lambda, \forall z' \in O^i \}, \\
    X_\lambda &:= \{ z \in X \mid z' \in X, \forall z' \text{ with } \|z - z'\| \leq \lambda \}.
\end{align*}

We now verify that for each $i = 1, \dots, N_s$, the system trajectory $x$ satisfies reachability $\eventually G^i$, avoidance $\always \neg(O^i)$, and invariance $\always X$.

\textbf{Reachability:} Since $\xi \models \eventually G_\lambda^i$, there exists $t_i \in \R_0^+$ such that $\xi(t_i) \in G_\lambda^i$. By definition of $G_\lambda^i$, this implies that the ball $\mathcal{B}(\xi(t_i), \lambda)$ is entirely contained in $G^i$. As the system trajectory remains within the VCZ ball, in particular $x(t_i) \in \mathcal{B}(\xi(t_i), \lambda)$, it follows that $x(t_i) \in G^i$. Hence, $x$ satisfies $\eventually G^i$.

\textbf{Avoidance:} Since $\xi \models \always \neg (O_\lambda^i)$, we have $\xi(t) \notin O_\lambda^i$, for all $t \in \R_0^+$, i.e., $\|\xi(t) - z' \| > \lambda,$ for all $z' \in O^i$. 
From triangle inequality, for all $t \in \R_0^+, z' \in O^i$,
\begin{align*}
    \| x(t) - z' \| & \geq \| \xi(t) - z' \| - \| x(t) - \xi(t) \| > \lambda - \lambda = 0,
\end{align*}
which shows that $x(t) \notin O^i$. Thus $x$ satisfies $\always \neg (O^i)$.

\textbf{Invariance:} Since $\xi \models \always X_\lambda$, we have $\xi(t) \in X_\lambda$ for all time $t \in \R_0^+$. By definition of $X_\lambda$, this means $\mathcal{B}(\xi(t), \lambda) \subseteq X$. As the system trajectory always lies within the VCZ ball, $x(t) \in \mathcal{B}(\xi(t), \lambda)$, it follows that $x(t) \in X$. Hence, $x$ satisfies $\always X$.

Since the above argument holds for each $i \in [1;N_s]$, it follows that $x$ satisfies all the RAS subtasks in $\phi$. Therefore, $x \models \phi$. \qed
\end{pf}

\section{Proof of Theorem \ref{thm:symbol}}\label{sec:symbol_proof}
\begin{pf}
The proof is carried out in two steps.

First, since $\Sigma_h(\V) \preceq_Q \Sigma_q(\V)$, the feedback refinement relation ensures that any controller designed using the symbolic model can be refined for the sampled system. Thus, if the abstract controller $C_q$ guarantees $\Sigma_q(\V) \mid C_q \models \hat{\phi}_{\lambda+\delta}$, the refined controller $C_h := C_q \circ Q$ guarantees $\Sigma_h(\V) \mid C_h \models \phi_{\lambda+\delta}$. This follows directly from \cite[Theorem VI.3]{Reissig2017feedback}.

Second, to extend correctness from the sampled system to the continuous system, we account for inter-sample behavior. Following \cite[Theorem 4.1]{liu2016finite}, modifying the specification by a robustness margin $\delta = \frac{\bar{u}h}{2} + \eta$, where $\eta$ is the quantization parameter, $\bar{u}$ is the input bound, and $h$ is the sampling-period, ensures that satisfaction of $\phi_{\lambda+\delta}$ by the sampled system implies satisfaction of $\phi_\lambda$ by the continuous system $\Sigma(\V)$.\qed
\end{pf}

\section{Proof of Theorem \ref{thm:mainresult}}\label{sec:cont_proof}
\begin{pf}
    The proof is divided into two stages:

    \subsection*{Stage I: Confinement within the VCZ.} We aim to show that the velocity command $v_r(t)$ defined in \eqref{eqn:velcon} ensures the system configuration $x(t)$ remains within the VCZ for all time $t \in \R_0^+$. 

    We proceed by contradiction. Suppose there exists a time $\tx\in \R^+$ such that $\tx$ is the first instance when $x(t)$, on the application of velocity input $v_r(t)$ \eqref{eqn:velcon} exits the VCZ. Then,
    \begin{gather}\label{Eq:inqe_tx}
        \|x(t) - \xi(t)\| < \lambda, \text{ for all } t \in [0, \tx).
    \end{gather}

    As the $x(t)$ approaches the VCZ boundary, $\|x(t) - \xi(t)\| \rightarrow \lambda$, for crossing the boundary, we have the following implications:     
    \begin{align*}
        &\|x(t) - \xi(t)\| < \lambda \implies \|x(t) - \xi(t)\| \uparrow \lambda \\
        \implies &\lim_{\|x(t) - \xi(t)\| \uparrow \lambda} \frac{d}{dt} \|x(t) - \xi(t)\| > 0 \\
        \implies &\lim_{\|x(t) - \xi(t)\| \uparrow \lambda} (x(t) - \xi(t))^\top (\dot{x}(t) - \dot{\xi}(t)) > 0 \\
        \implies &\lim_{\|x(t) - \xi(t)\| \uparrow \lambda} (x(t) - \xi(t))^\top \dot{x}(t) \\
        & \qquad \qquad \ > \lim_{\|x(t) - \xi(t)\| \uparrow \lambda}(x(t) - \xi(t))^\top \dot{\xi}(t).
    \end{align*}
    Substituting the dynamics from Equations \eqref{eqn:vcz} and \eqref{eqn:sysDyn_vel},
    \begin{align*}
        &\lim_{\|x(t) - \xi(t)\| \uparrow \lambda} (x(t) - \xi(t))^\top v \\
        & \qquad \qquad \ > \lim_{\|x(t) - \xi(t)\| \uparrow \lambda}(x(t) - \xi(t))^\top u.
    \end{align*}    
    Using the velocity command $v_r$ from \eqref{eqn:velcon} and the fact that $\lim_{e_x \rightarrow 1} \Psi(e_x) = 1$,
    \begin{align*}
        &\lim_{\|x(t) - \xi(t)\| \uparrow \lambda} -\bar{v} (x(t) - \xi(t))^\top \frac{(x(t) - \xi(t))}{\|x(t) - \xi(t)\|} \\
        & \qquad \qquad \qquad \qquad > \lim_{\|x(t) - \xi(t)\| \uparrow \lambda}(x(t) - \xi(t))^\top u \\
        \implies &\lim_{\|x(t) - \xi(t)\| \uparrow \lambda} \bar{v} \|x(t) - \xi(t)\| \\
        & \qquad \qquad \qquad \qquad < \lim_{\|x(t) - \xi(t)\| \uparrow \lambda} -(x(t) - \xi(t))^\top u \\
        \implies &\bar{v} \lambda < \lim_{\|x(t) - \xi(t)\| \uparrow \lambda} \|-(x(t) - \xi(t))^\top \| \| u \| \leq \lambda \bar{u} \\
        \implies &\bar{v} < \bar{u}.
    \end{align*}
    However, this contradicts the feasibility constraint in Equation \eqref{eqn:feas}. Hence, $\|x(t) - \xi(t)\| \nrightarrow \lambda, \forall t \in [0,\tx)$, i.e., $x(t)$ never approaches the VCZ boundary over $t \in [0,\tx)$.
    Consequently, due to the continuity of $x(t)$, it can be concluded that there is no $\tx$ at which $x(t)$ crosses the VCZ boundary.
    
    Therefore, given a reference velocity vector \eqref{eqn:velcon},
    \begin{gather*}
        \|x(t) - \xi(t)\| < \lambda, \forall t \in \R^+_0.
    \end{gather*}

    \subsection*{Stage II: Tracking of reference velocity.} 
    Next, we show that the control input $\tau(t)$ defined in \eqref{eqn:con} ensures the velocity error $e_v(t)$ remains within the funnel for all time $t \in \R_0^+$ as in \eqref{eqn:fun2}. 

    We proceed by contradiction. Let $\tx$ be the first time instance when the velocity error $e_v(t)$, on the application of input $\tau(t)$ \eqref{eqn:con}, violates \eqref{eqn:fun2}, 
    $$\exists i \in [1;n], e_{v,i}(\tx) \leq -\rho_{v,i}(\tx) \text{ or } e_{v,i}(\tx) \geq \rho_{v,i}(\tx).$$
    Then,
    \begin{gather}\label{Eq:inqe_tv}
        -\rho_{v,i}(t) < e_{v,i}(t) < \rho_{v,i}(t), \forall (t,i) \in [0, \tx) \times [1;n].
    \end{gather}
    We examine the following two cases for $t \in [0,\tx)$.

\textbf{Case I.} There exists $i \in [1;n]$ such that $e_{v,i}(t)$ approaches the upper funnel constraint, i.e., $e_{v,i}(t) \rightarrow \rho_{v,i}(t) \implies e_{v,i}(t) - \rho_{v,i}(t) \rightarrow 0$. Following \eqref{Eq:inqe_tv}, we have the following implications:
\begin{align*}
    &e_{v,i}(t) < \rho_{v,i}(t) \implies (e_{v,i}(t) - \rho_{v,i}(t)) \uparrow 0 \\
    \implies &\lim_{(e_{v,i}(t) - \rho_{v,i}(t)) \uparrow 0} \frac{d}{dt} (e_{v,i}(t) - \rho_{v,i}(t)) > 0 \\
    \implies &\lim_{(e_{v,i}(t) - \rho_{v,i}(t)) \uparrow 0} \dot{e}_{v,i}(t) > \lim_{(e_{v,i}(t) - \rho_{v,i}(t)) \uparrow 0} \dot{\rho}_{v,i}(t) \\
    \implies &\lim_{(e_{v,i}(t) - \rho_{v,i}(t)) \uparrow 0} \dot{e}_{v,i}(t) > -\mu_{v,i}(p_{v,i}-q_{v,i})e^{-\mu_{v,i}t} \\
    \implies &\lim_{(e_{v,i}(t) - \rho_{v,i}(t)) \uparrow 0} \dot{e}_{v,i}(t) > -\mu_{v,i}(p_{v,i}-q_{v,i}) \\
    \implies &\lim_{(e_{v,i}(t) - \rho_{v,i}(t)) \uparrow 0} \dot{v}_{i}(t) > -\mu_{v,i}(p_{v,i}-q_{v,i}) + \dot{v}_r(t) \\
    \implies &\lim_{(e_{v,i}(t) - \rho_{v,i}(t)) \uparrow 0} \dot{v}_{i}(t) > -\mu_{v,i}(p_{v,i}-q_{v,i}) - \overline{a}_{r,i}.
\end{align*}
Therefore, there exists $i \in [1;n]$, such that 
\begin{gather}\label{eqn:dv_b1}
    \lim_{(e_{v,i}(t) - \rho_{v,i}(t)) \uparrow 0} \dot{v}_{i}(t) > -\mu_{v,i}(p_{v,i}-q_{v,i}) - \overline{a}_{r,i}.
\end{gather}

Now, let us look at the control input vector $\tau(t) = [\tau_{1}(t), \ldots, \tau_{n}(t)]^\top$, for all $i \in [1;n]$
\begin{align*}
    &\lim_{(e_{v,i}(t) - \rho_{v,i}(t)) \uparrow 0} \varepsilon_{v,i}(t) = 1, \\
    \implies &\lim_{(e_{v,i}(t) - \rho_{v,i}(t)) \uparrow 0} \tau_{i}(t) = -\overline{\tau}_{i}.
\end{align*}
Given the acceleration-level system dynamics \eqref{eqn:sysDyn_acc} and feasibility constraints \eqref{eqn:feas}
\begin{align}\label{eqn:dv_b1c}
    &\lim_{(e_{v,i}(t) - \rho_{v,i}(t)) \uparrow 0} \dot{v}_{i}(t) \leq \overline{V}_{M,i} - \underline{m} \overline{\tau}_i + \underline{m}_i\overline{d}_i, \nonumber \\
    \implies &\lim_{(e_{v,i}(t) - \rho_{v,i}(t)) \uparrow 0} \dot{v}_{i}(t) \leq -\mu_{v,i}(p_{v,i}-q_{v,i}) - \overline{a}_{r,i}.
\end{align}
Thus, \eqref{eqn:dv_b1} contradicts \eqref{eqn:dv_b1c}. Hence, $e_{v,i}(t) \nrightarrow \rho_{v,i}(t), \forall (t,i) \in [0,\tx) \times [1;n]$, i.e., the velocity error $e_v(t)$ never approaches the upper funnel constraint over $t \in [0,\tx)$ in any dimension.

\textbf{Case II.} There exists $i \in [1;n]$ such that $e_{v,i}(t)$ approaches the lower funnel constraint, i.e., $e_{v,i}(t) \rightarrow -\rho_{v,i}(t) \implies e_{v,i}(t)+\rho_{v,i}(t) \rightarrow 0$. Following \eqref{Eq:inqe_tx}, we have the following implications:
\begin{align*}
    &e_{v,i}(t) > -\rho_{v,i}(t) \implies (e_{v,i}(t)+\rho_{v,i}(t)) \downarrow 0 \\
    \implies &\lim_{(e_{v,i}(t)+\rho_{v,i}(t)) \downarrow 0} \frac{d}{dt} (e_{v,i}(t)+\rho_{v,i}(t)) < 0 \\
    \implies &\lim_{(e_{v,i}(t)+\rho_{v,i}(t)) \downarrow 0} \dot{e}_{v,i}(t) < \lim_{(e_{v,i}(t)+\rho_{v,i}(t)) \downarrow 0} -\dot{\rho}_{v,i}(t) \\
    \implies &\lim_{(e_{v,i}(t) + \rho_{v,i}(t)) \downarrow 0} \dot{e}_{v,i}(t) < \mu_{v,i}(p_{v,i}-q_{v,i})e^{-\mu_{v,i}t} \\
    \implies &\lim_{(e_{v,i}(t) + \rho_{v,i}(t)) \downarrow 0} \dot{e}_{v,i}(t) < \mu_{v,i}(p_{v,i}-q_{v,i}) \\
    \implies &\lim_{(e_{v,i}(t) + \rho_{v,i}(t)) \uparrow 0} \dot{v}_{i}(t) < \mu_{v,i}(p_{v,i}-q_{v,i}) + \dot{v}_r(t) \\
    \implies &\lim_{(e_{v,i}(t) + \rho_{v,i}(t)) \uparrow 0} \dot{v}_{i}(t) < \mu_{v,i}(p_{v,i}-q_{v,i}) + \overline{a}_{r,i}.
\end{align*}
Therefore, there exists $i \in [1;n]$, such that
\begin{gather}\label{eqn:dv_b2}
    \lim_{(e_{v,i}(t) + \rho_{v,i}(t)) \uparrow 0} \dot{v}_{i}(t) < \mu_{v,i}(p_{v,i}-q_{v,i}) + \overline{a}_{r,i}.
\end{gather}

Now, let us look at the control input vector $\tau(t) = [\tau_{1}(t), \ldots, \tau_{n}(t)]^\top$, for all $i \in [1;n]$
\begin{align*}
    &\lim_{(e_{v,i}(t)+\rho_{v,i}(t)) \downarrow 0} \varepsilon_{v,i}(t) = -1, \\
    \implies &\lim_{(e_{v,i}(t)+\rho_{v,i}(t)) \downarrow 0} \tau_{i}(t) = \overline{\tau}_i.
\end{align*}
Given the acceleration-level system dynamics \eqref{eqn:sysDyn_acc} and feasibility constraints \eqref{eqn:feas}
\begin{align}\label{eqn:dv_b2c}
    &\lim_{(e_{v,i}(t)+\rho_{v,i}(t)) \downarrow 0} \dot{v}_{i}(t) \geq \underline{V}_{M,i} + \underline{m} \overline{\tau}_i - \underline{m}_i\overline{d}_i, \nonumber \\
    \implies &\lim_{(e_{v,i}(t)+\rho_{v,i}(t)) \downarrow 0} \dot{v}_{i}(t) \geq \mu_{v,i}(p_{v,i}-q_{v,i}) + \overline{a}_{r,i}.
\end{align}
Thus, \eqref{eqn:dv_b2} contradicts \eqref{eqn:dv_b2c}. Hence, $e_{v,i}(t) \nrightarrow -\rho_{v,i}(t), \forall (t,i) \in [0,\tx) \times [1;n]$, i.e., the velocity error $e_v(t)$ never approaches the lower funnel constraint over $t \in [0,\tx)$ in any dimension.

Thus, over $t \in [0, \tx])$, $e_{v,i}(t)$ never approaches the funnel constraints $-\rho_{v,i}(t)$ and $\rho_{v,i}(t)$ for all $i \in [1;n]$.
Consequently, due to the continuity of $e_v(t)$, it can be concluded that there is no $\tx$ at which $e_{v,i}(t)$ violates the funnel constraints $-\rho_{v,i}(t)$ and $\rho_{v,i}(t)$ for all $i \in [1;n]$.
Therefore, given input \eqref{eqn:con},
\begin{gather*}
    -\rho_{v,i}(t) < e_{v,i}(t) < \rho_{v,i}(t), \forall (t,i) \in \R^+_0 \times [1;n].
\end{gather*}


Hence, the bounded control input $\tau(t)$ in \eqref{eqn:con}, under feasibility conditions \eqref{eqn:feas}, ensures confinement of the system configuration $x$ within the VCZ $\B(\xi, t)$. \qed
\end{pf}

\end{document}